\documentclass[a4paper,10pt]{article}
\usepackage{color}
\usepackage{hyperref} 
 \usepackage{latexsym}
  \usepackage{amssymb}
  \usepackage{graphicx}
\usepackage[utf8]{inputenc}
\usepackage[T1]{fontenc}
 \usepackage[english]{babel}
  \usepackage{}
  \usepackage{amstext}
  \usepackage{amsmath}
  \usepackage{times} 
  \usepackage{amsfonts}
 \date{}

\title{ A Study of  the forced Van der Pol generalized oscillator  with renormalization group method.}
\author{L. A. Hinvi \footnote{laurent.hinvi@imsp-uac.org}\hspace{0.08cm} 
A. V. Monwanou\footnote{movins$2008$@yahoo.fr} \hspace{0.05cm}  and J. B. Chabi
   Orou\footnote{Author to whom correspondence should be addressed: jchabi@yahoo.fr}}

\begin{document}

\maketitle {Institut de Math\'ematiques et de Sciences Physiques, BP: 613 Porto- Novo, B\'enin}

\begin{abstract}
In this paper the equation of forced Van der Pol generalized oscillator is examined with renormalization group
method. A brief recall of the renormalization group technique is done. We have applied this method to the
equation of forced Van der Pol generalized oscillator to search for its asymptotic solution and its renormalization
group equation. The analysis of the numerical simulation graph is done; the method’s efficiency is pointed out.
\end{abstract}
\section{Introduction}
The analysis of the asymptotic behavior has played an important role in  applied mathematics
and theoretical physics. In many cases, the regular perturbation methods
become inapplicable  than the  singular perturbation  methods see (Bender and Orszag, $1978$ ;
Chen, Goldenfeld, and Oono, $1996$; Chiba, $2008b$; Hinch, $1991$) \cite{1} - \cite{4}.  
We can cite as singular perturbation methods  for solving ordinary 
differential equations $(ODE)$, methods multiple scales,  $WKB$ (Bender and Orszag, $1978$)
\cite{1}, the method recovery (Roberts, $1985$) \cite{3},
 etc. The renormalization group method  which is the subject of this study was compiled
 by Chen, Goldenfeld, Oono ($1994,1996$) 
\cite{4} for differential equations form
 \begin{eqnarray}
\dot{x} &=&F x+ \varepsilon g(t,x,\varepsilon) ;\quad  x\in \mathbb{R}^n \label{eq1}
\end{eqnarray}
where $\varepsilon \geq 0$ is a  small parameter.
They showed that the  Renormalization group method  unifies the singular perturbation methods 
listed above  \cite{4}.
With this method, the renormalization constants of integration can raise divergence.
This technique of renormalization  does appear the renormalization group equation $(RGE)$  of
involving the amplitude which stabilizes  the limit cycle; it is simple for dynamical system
analysis. Chiba ($2008b$) used the renormalization group method to analyze the model of Kuramoto
coupled oscillators.

The Van der Pol equation is a basic model for oscillatory processes in physics,
electronics, biology, neurology, sociology and economic (Marios, $2006$). 
In this work we decided to investigate the forced Van der Pol oscillator in its 
generalized form governed by the dimensionless equation below. We have done
the similar work  where the  unforced Van der Pol  generalized  oscillator is studied.
We will forced the system with a periodic extern force of pulsation $\Omega$. In this case 
$g(t,x,\varepsilon)$ is an explicit function of time. This oscillator has been applied for modeling
a Bipedal Robot by Pina Filho and Dutra ($2009$) and known as Hybrid 
Van der Pol-Rayleigh oscillators. Sarkar and Bhattacharjee ($2010$) recently studied the unforced
 Van der Pol oscillator with another technique of the renormalization group theory to find 
 its limit cycle.  The paper is organized as follows.
 
In the second section a brief recall   of the  renormalization group technique will be  done.
In the third section,  the method will be applied to the equation of forced Van der Pol
generalized oscillator. In the fourth section our results will be analyzed through  the graphs. 
The conclusions will be presented in the
final section.

\section{Renormalization group method}

In this section we recall the outline of the technical group renormalization for $(ODE)$.
For more details we refer to (chiba, $2008b$)\cite{1}. We consider an ODE of the form :
 \begin{eqnarray}
\dot{x} &=&Fx+\varepsilon g(t,x,\varepsilon)\nonumber\\
&=& F x+\varepsilon g_{1}(t,x)+\varepsilon^2 g_{2}(t,x)+\dots;
x\in \mathbb{R}^n, \label{eq2}
\end{eqnarray}
where $0\leq \varepsilon \leq1 $. For this system, we assume that:
\begin{enumerate}
\item The matrix $F$ is a diagonalizable $n\times n$  constant matrix all
of whose eigenvalues lie on the imaginary axis.
\item The function $g(t, x, \varepsilon)$ is  of $C^{\infty}$ class with 
respect to $t$, $x$ and $\varepsilon$.
The formal power series expansion of $g(t, x, \varepsilon)$  in $\varepsilon$ 
is given as above.
 \item each $g_{i} (t, x)$ is periodic in  $t\in \mathbb{R}$ and polynomial in $x$.
\end{enumerate}
Firstly we apply the simple development method and secondly the
renormalization group method will be applied to break down the divergence.
We replace $x$ in  Equation (\ref{eq2}) by
\begin{equation}
  x(t)=x_0(t)+\varepsilon x_1(t)+\varepsilon^2 x_{2}(t)+\dots \label{eq3}
\end{equation}
After development and identification of the coefficients of $\varepsilon$ 
we find:

\begin{equation}
 \dot{x}_{0}= Fx_0 ,  \label{eq4}
\end{equation}
\begin{equation}
\dot{x}_{i} = Fx_{i}+G_{i}(t,x_{0},x_{1},x_{i-1});  \label{eq6}
\end{equation}
where the homogeneous term $G_{i}$ is a regular function of 
$t$, $x_{0}$, $x_{i-1}$ with:

 \begin{eqnarray}
 G_{1}(t,x_{0} )  &=&  g_{1}(t,x_{0}),\\  \label{eq7}
G_{2}(t,x_{0},x_{1})  &=&  \frac{{\partial}{g_{1}}}{{\partial}{x}}(t,x_{0}) x_{1} +g_{2}(t,x_{0}) ,    \label{eq8}
\end{eqnarray}
 We can verify the Equality (see   Chiba, $2008a$,{\bf{lemma A.2 for the proof}} \cite{7} ) :
 \begin{equation}
 \frac{{\partial}{G_{i}}}{{\partial}{x_{j}}}=\frac{{\partial}{g_{i-1}}}{{\partial}{x_{j-1}}}=
 \frac{{\partial}{g_{i-j}}}{{\partial}{x_{0}}}  ,i> j\geq0 .
 \label{eq11} 
 \end{equation}

 In what follows, we denote the fundamental matrix $e^{Ft}$ as $X(t)$. Define the functions 
 $R_{j}$, $h_{t}^{(i)}$,  $i=1,2,$ $\cdots$
 on $\mathbb {R}^n$ by

 \begin{eqnarray}
 R_{1}(y)&=& \displaystyle \lim_{t \rightarrow +\infty}  {\frac{1}{t} 
 \int_{t_{0}}^{t}[ X(s)^{-1}G_{1}(s,X(s)y)] ds} \label{eq12}\\
{h}^{1}_{t}(y)& = & X(t)\int_{t_{0}}^{t}[ X(s)^{-1}G_{1}(s,X(s)y)- R_{1}(y)] ds \label{eq13}
\end{eqnarray}
\begin{eqnarray}
 R_{i}(y)  &=& \displaystyle \lim_{t \rightarrow +\infty}  \frac{1}{t} \int_{t_{0}}^{t}
 [ X(s)^{-1}G_{i}(s,X(s)y,h^{1}_{s}(y),\dots,h^{i-1}_{s}(y)) \cr
 &&- X(s)^{-1}\sum_{k=1}^{i-1} (D{h^{k}_{s})_{y} R_{i-k}(y)] ds} \label{eq14}
,i=2,3\cdots   \\
 {h}^{i}_{t}(y) & =& X(t)\int_{t_{0}}^{t}[ X(s)^{-1}G_{i}(s,X(s)y,h^{1}_{s}(y),\dots,h^{i-1}_{s}(y)) \cr
 &&- X(s)^{-1}\sum_{k=1}^{i-1}(D{h^{k}_{s})_{y} R_{i-k}(y)-R_{i}(y] ds} .\label{eq15}
\end{eqnarray}
 
 {\bf{ Proposal}} Let $x_{0}(t)=X(t)y$  be the solution to  (\ref{eq4})
 whose initial value is 
$y\in \mathbb{R}^{n}$. Then for an arbitrary time $\tau \in \mathbb{R}$  
and $i=1,2,3 \dots$, the curve defined  $x_{i}$ 
by:
 \begin{eqnarray}
 x_{i}& :=&x_{i}(t,\tau,y)\nonumber\\
&=&{h}^{i}_{t}(y) + p_{1}^{i}(t,y)\left(t-\tau\right) + p_{2}^{i}(t,y)
\left(t-\tau\right)^{2} +\dots+ p_{i}^{i}(t,y)\left(t-\tau\right)^{i} ; 
  \label{eq16}
\end{eqnarray}
 gives a solution to Equation  (\ref{eq6}), where the functions $p_{1}^{i},\dots, p_{i}^{j}$ are given by :
 \begin{eqnarray}
 p_{1}^{i}(t,y) & =&  X(t)R_{i}(y)+\sum_{k=1}^{i-1}(D{h^{k}}_{t})_{y} R_{i-k}(y),  \label{eq17}\\
p_{j}^{i}(t,y) & =&  \frac{1}{j}\sum_{k=1}^{i-1} \frac{{\partial}{ p_{j-1}^{k}}} 
{\partial y}(t,y)R_{i-k}(y),(j=2,3,\dots,i-1),  \label{eq18}
\end{eqnarray}
 Further, the functions $h_{t}^{(i)}$ are bounded uniformly in $t$.
The solution of the  Equation (\ref{eq2})  is given by :
 \begin{eqnarray}
 x(t,\tau,y)&=&x_{0}+\varepsilon x_{1} \nonumber\\ 
 &=&X(t)y+\varepsilon(h_{t}^{1}(y)+X(t)R_{1}(y)(t-\tau))+ O(\varepsilon^{2}).\label{eq21} 
\end{eqnarray}
 It is the solution obtained by simple development , it diverges for time long, leading to
 the need for its 
  renormalization. It should not depend on  $\tau$  
  $(\frac{\partial{x(t,\tau,y(\tau))}}{\partial{\tau}}|_{\tau=t}=0)$, then 
\begin{eqnarray}
  0=X(t)\frac{dy(t)}{dt}+\varepsilon \frac{\partial{h_{t}^{1}}}{\partial{y}} 
  \frac{d{y(t)}}{d{t}}-\varepsilon X(t)R_{1}(y). \label{eq22}
 \end{eqnarray}
  We verify that (\ref{eq22}) admits solution :
\begin{eqnarray}
 \frac{d{y(t)}}{d{t}}&=& \varepsilon R_{1}(y)+O(\varepsilon^{2}). \label{eq23}
\end{eqnarray}
Let $y (t)$ be a solution of Equation (\ref{eq23}), then the solution of Equation 
(\ref{eq2}) looked for the
renormalization  group method is given by :
 
 \begin{eqnarray}
   x(t,t,y)=X(t)y(t)+\varepsilon h_{t}^{1}\left(y(t)\right)  +O(\varepsilon^{2}). \label{eq24}
 \end{eqnarray}
 The Equation (\ref{eq23}) is the equation of the renormalization group of Equation (\ref{eq2}). 
 The calculation for a higher order is in the same way and one obtains the 
 equation of renormalization group of order $m$ as follows :
 \begin{equation}
 \frac{dy}{dt}=\varepsilon R_{1}(y)+\varepsilon^{2}R_{2}(y)+\dots +\varepsilon^{m}R_{m}(y), 
 y\in \mathbb{R}^n.
 \end{equation}

 \section{Application  to the  Forced Van der Pol  Generalized oscillator}
 We consider the forced Van der Pol  generalized oscillator gouverning by 
 dimensionless equation as follows
 \begin{eqnarray}
\ddot{x} + x-\varepsilon(1-ax^{2}-b\dot{x}^{2})\dot{x}=E\sin{\Omega t}; \label{eq27}
\end{eqnarray}
where $a$,  $b$,  $\varepsilon$, $E$ and  $\Omega$ are positifs control parameters such as
$\varepsilon$ is small. $E\sin{\Omega t}$ is extern force for pulsation $\Omega$ and amplitude
$E$. The internal pulsation is here equal to one. 
With $E=\varepsilon c$,  $y=\dot{x}$, $x=(z+\bar{z})$, $y=i(z-\bar{z})$
we rewrite  Equation (\ref{eq27})  as

\begin{eqnarray}
\left\{
 \begin{array}{cl}
\dot{ z}= iz+   \frac{ \varepsilon}{2}[(z- \bar{z})-a(z+\bar{z})^{2}(z-\bar{z})
+b(z-\bar{z})^{3} -ic\sin{\Omega t}]                     
\\
\\
\dot{\bar{z}}=-i\bar{z}+\frac{\varepsilon}{2}[-(z-\bar{z})+a(z+\bar{z})^{2}
(z-\bar{z})-b(z-\bar{z})^{3}+ic\sin{\Omega t} ] \label{eq28}
\end{array}
\right
.\end{eqnarray}

The two equations of the system are nearly identical, 
the problem amounts to solving one of them. with

\begin{equation}
 z=z_{0}+\varepsilon z_{1}+\varepsilon^{2} z_{2} +\dots \label{eq29}
\end{equation}
we find
\begin{eqnarray} 
 \dot{z_{0}} &= & iz_{0}  \label{eq30}, \\      
\dot{z_{1}} &=& iz_{1}+ G_{1}(t,z_{0}). \label{eq31}
\end{eqnarray}
From zero order we have :
\begin{equation}
z_{0}=qe^{it} = qZ(t)\label{eq32}
\end{equation}

with $q$ the integration constant of  Equation (\ref{eq30}).
Expressions (\ref{eq12}) and (\ref{eq13})  give
\begin{eqnarray}
 R_{1}(q) &=& \frac{1}{2}q(1-(a+3b)|q|^{2})-{ic}
\displaystyle \lim_{t \rightarrow +\infty}\frac{1}{2t} \int_{t_{0}}^{t} Z(s)^{-1}
\sin({\Omega s})ds,  \label{eq33}
\end{eqnarray}
\begin{eqnarray}
h_{t}^{1}(q)&=& \frac{i}{4} [ (a-b)( q^{3}e^{3it}+\frac{1}{2}\bar{q}^{3}e^{-3it}) 
+((a+3b)q\bar{q}^{2}-\bar{q}) e^{-it}]\cr
&&-\frac{ic}{2}e^{it}
\int_{t_{0}}^{t} Z(s)^{-1}\sin({\Omega s})ds. \label{eq34}
\end{eqnarray}
where $t_{0}$ is an initial time and $Z(s)=e^{is}$
We find after computation:
\begin{eqnarray}
 R_{1}(q) &=& \frac{1}{2}q(1-(a+3b)|q|^{2}), \label{eq35}\\
 h_{t}^{1}(q)&=& \frac{i}{4} [ (a-b)( q^{3}e^{3it}+\frac{1}{2}\bar{q}^{3}e^{-3it}) 
+((a+3b)q\bar{q}^{2}-\bar{q}) e^{-it}]\cr \label{eq36}
&&-\frac{ic}{2}e^{it}I,
\end{eqnarray}
with
\begin{eqnarray}\label{37}
 I &=&\frac{1}{2} \left(\frac{\cos(1-\Omega)t}{1-\Omega}-
 \frac{\cos(1+\Omega)t}{1+\Omega} \right)-\cr
&&\frac{i}{2}\left(\frac{\sin(1-\Omega)t}{1-\Omega}-
\frac{\sin(1+\Omega)t}{1+\Omega} \right) ; 1\neq \Omega.
\end{eqnarray}

According to the proposal and the above results we have :

\begin{eqnarray}
 z(t,\tau,q)=qZ(t)+\varepsilon (h_{t}^{1}(q) +R_{1}(q)(t-\tau)) +O(\varepsilon^{2}),
 \label{eq38}
\end{eqnarray}
which diverges for  long $t$  because of the last term.
Using the notion of renormalization constant of integration
$({\frac{\partial{x(t, \tau,q)}}{\partial{\tau}}}| _{\tau=t} =0)$ 
mentioned in the previous section 
and taking $q(\tau)=r(\tau)e^{i\Theta(\tau)}$ we find :
\begin{eqnarray}\label{eq39}
x(r,t,\Theta) &=&2r\cos(t+\Theta (\tau))- \frac{r\varepsilon}{2} \sin(t+\Theta(\tau) )\cr
&&+ \frac{\varepsilon}{2} \left(\frac{(b-a)}{2}r^{3}\sin(3t+3\Theta (\tau))  + 
(a+3b)r^{3}\sin(t+\Theta(\tau)) \right)+\cr
&&+\frac{E }{(1-\Omega^{2})}\sin{\Omega t} +O(\varepsilon^{2}). 
\end{eqnarray}

\begin{eqnarray}
\left\{
 \begin{array}{cl}
\frac{dr}{d\tau}=  \frac{\varepsilon r}{2}(1-(a+3b)r^{2})+O(\varepsilon^{2})   \label{eq40}                  
\\
\frac{d\Theta(\tau)}{d\tau}=0+O(\varepsilon^{2}).   
\end{array}
\right
.\end{eqnarray}

The first equation of system (\ref{eq40}) gives the stable cycle limit radius  
$r_{s}=\frac{1}{\sqrt{(a+3b)}}$, with $(a+3b)\neq0$.

For $a=0, b=\frac{1}{3}$ we have   Rayleigh forced oscillator equation and 
Equation (\ref{eq39}) takes the form 
 
\begin{eqnarray}
x(r=1,t,\Theta) &=&2\cos(t+\Theta)+\frac{E\sin{\Omega t}}{(1-\Omega^{2})}
+ \frac{\varepsilon}{12} \sin{3(t+\Theta)} 
+O(\varepsilon^{2}).\label{eq41}
\end{eqnarray} 
Also for $a=1, b=0$ we  have  the forced  Van der Pol oscillator and Equation
(\ref{eq39}) becomes
\begin{eqnarray}
x(r=1,t,\Theta) &=&2\cos(t+\Theta)+\frac{E\sin{\Omega t}}{(1-\Omega^{2})} -
\frac{1}{4}\varepsilon \sin{3(t+\Theta)}
 +O(\varepsilon^{2}).\label{eq42}
\end{eqnarray}
When we cancel the external force ($E=0$), the Equation (\ref{eq42}) reduces to 
the results found by Hasan($1981$) and recently by Sarkar and Bhattacharjee 
($2010$).

Finally for $a=1 ,b=1$  we  have  the forced  Van der Pol    generalized oscillator
equation and the    Equation (\ref{eq39}) becomes

\begin{eqnarray}
x(r=\frac{1}{2},t,\Theta) &=&\cos(t+\Theta)+\frac{E\sin{\Omega t}}{(1-\Omega^{2})}
 +O(\varepsilon^{2}).\label{eq43}
\end{eqnarray}

 The integration of the Equations system (\ref{eq40}) gives us:

\begin{eqnarray}
\left\{
 \begin{array}{cl}
r(t)= \frac{r_{0}e^{\varepsilon \frac{t}{2}}}{\sqrt{1+r_{0}^{2}(a+3b)(1-e^{\varepsilon t})}} 
+O(\varepsilon^{2}{t})\label{eq44}                  
\\
\\
\Theta(t)=\Theta_{0}+O(\varepsilon^{2}{t}).   
\end{array}
\right
.\end{eqnarray}

For $\varepsilon=0$ we have a  cycle limit radius $r$:

\begin{eqnarray}
\left\{
 \begin{array}{cl}
r(t)= r = \frac{r_{0}}{\sqrt{1+r_{0}^{2}(a+3b)}} 
\label{eq45}                  
\\
\\
\Theta(t)=\Theta_{0},   
\end{array}
\right
.\end{eqnarray}
It becomes, for $r_{0}=1$
\begin{eqnarray}
\left\{
 \begin{array}{cl}
r(t)= r = \frac{1}{\sqrt{1+(a+3b)}} 
\label{eq46}                  
\\
\\
\Theta(t)=\Theta_{0}.  
\end{array}
\right
.\end{eqnarray}

\section{ Analysis Results}
In this section we will do   through the graphical analysis of the numerical simulation 
figures below. These graphs are obtained on one hand  by  direct simulation of Equation
(\ref{eq27}) 
for   some parameters values and  on the other hand by simulation of the solution asymptotic 
Equation  (\ref{eq39}) found by the renormalization group method, for the same values of 
these parameters with the logician MATHEMATICA.

\begin{figure}[htbp]
 \begin{center}
 \includegraphics[width=3cm]{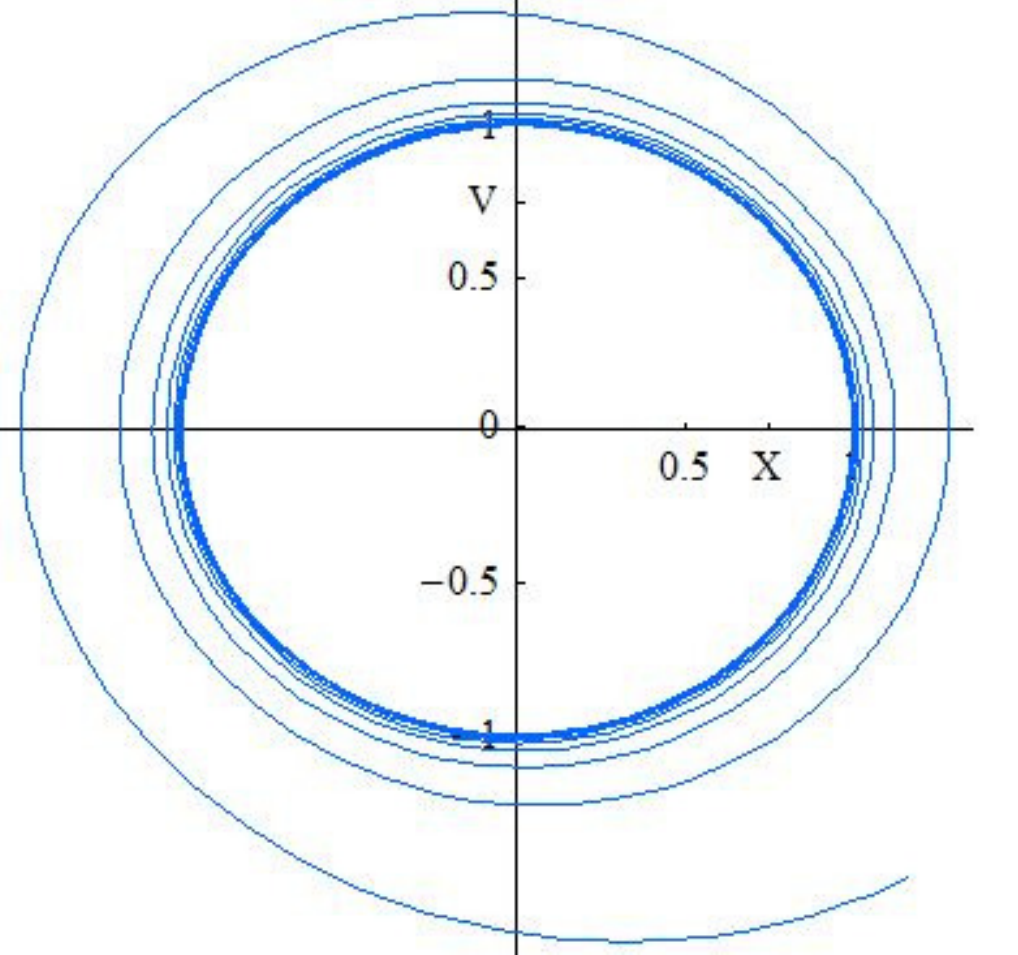}\includegraphics[width=3cm]{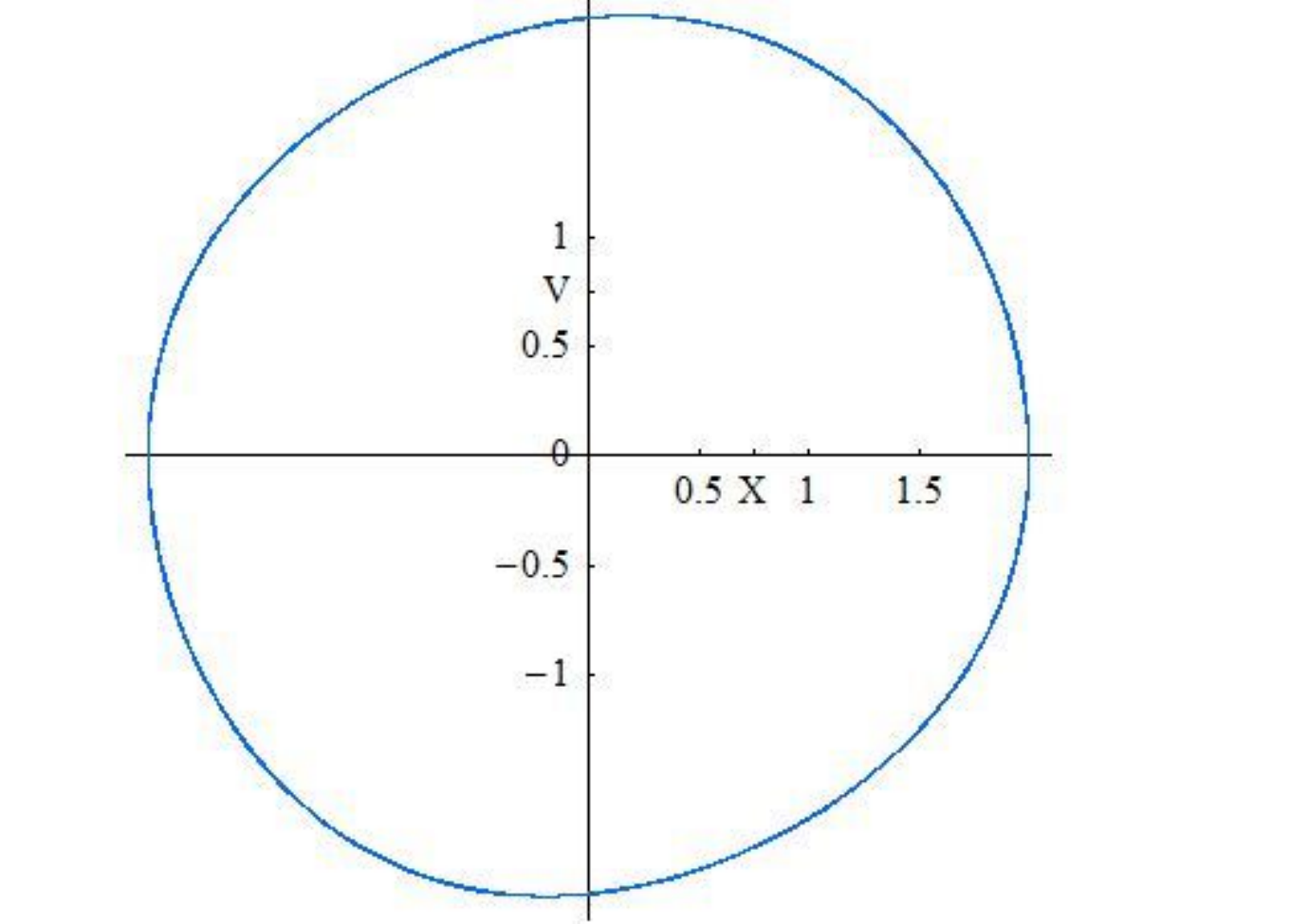}
 \includegraphics[width=3cm]{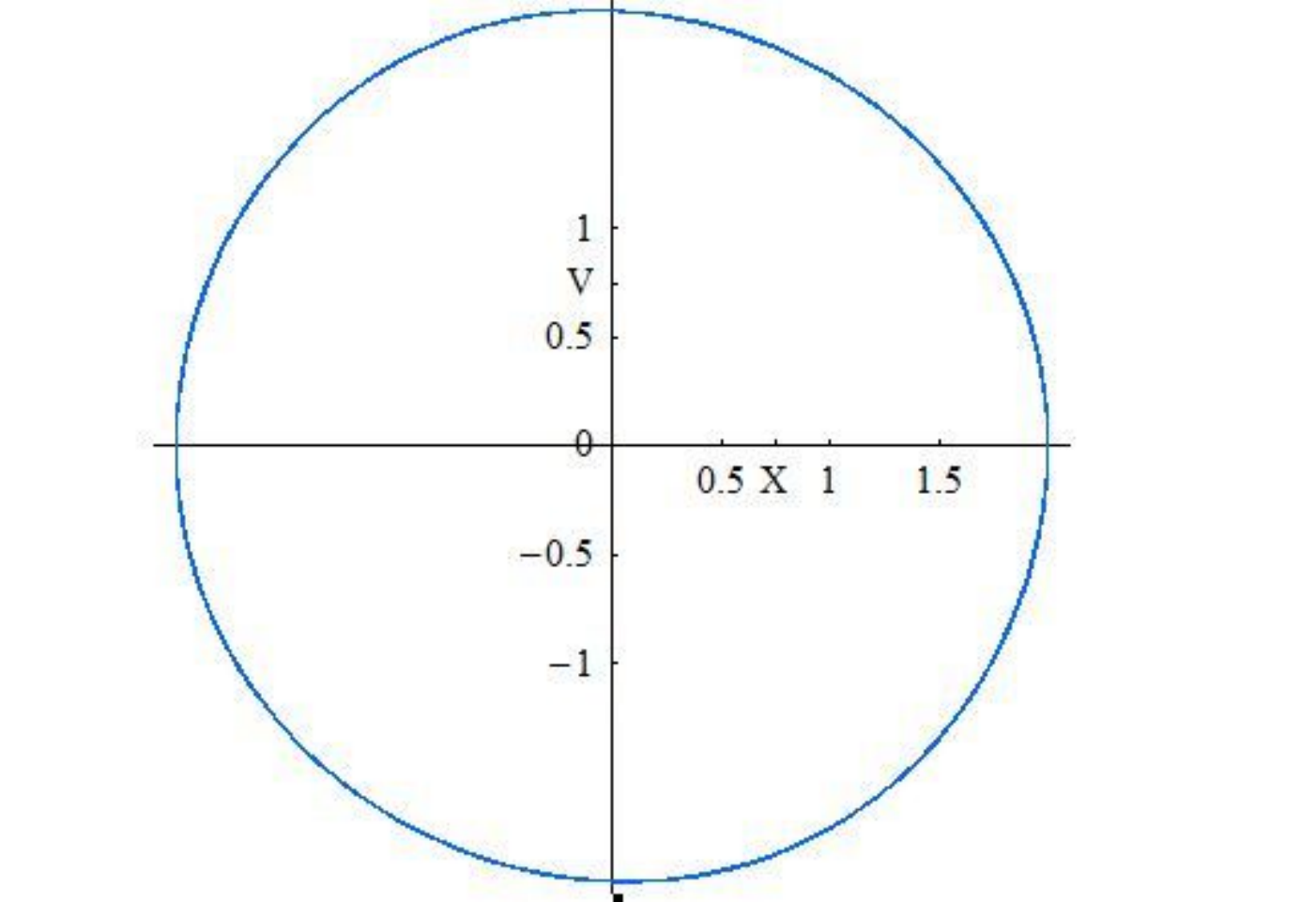} 
 \end{center}
 \caption{ phase diagram : Van der Pol generalized oscillator, Van der Pol  oscillator and   Rayleigh  oscillator,
 for $\varepsilon=10^{-1} , E=0$, $\Omega=0$.} 
 \label{fig:1}
 \end{figure}

 \begin{figure}[htbp]
  \begin{center}
  \includegraphics[width=3cm]{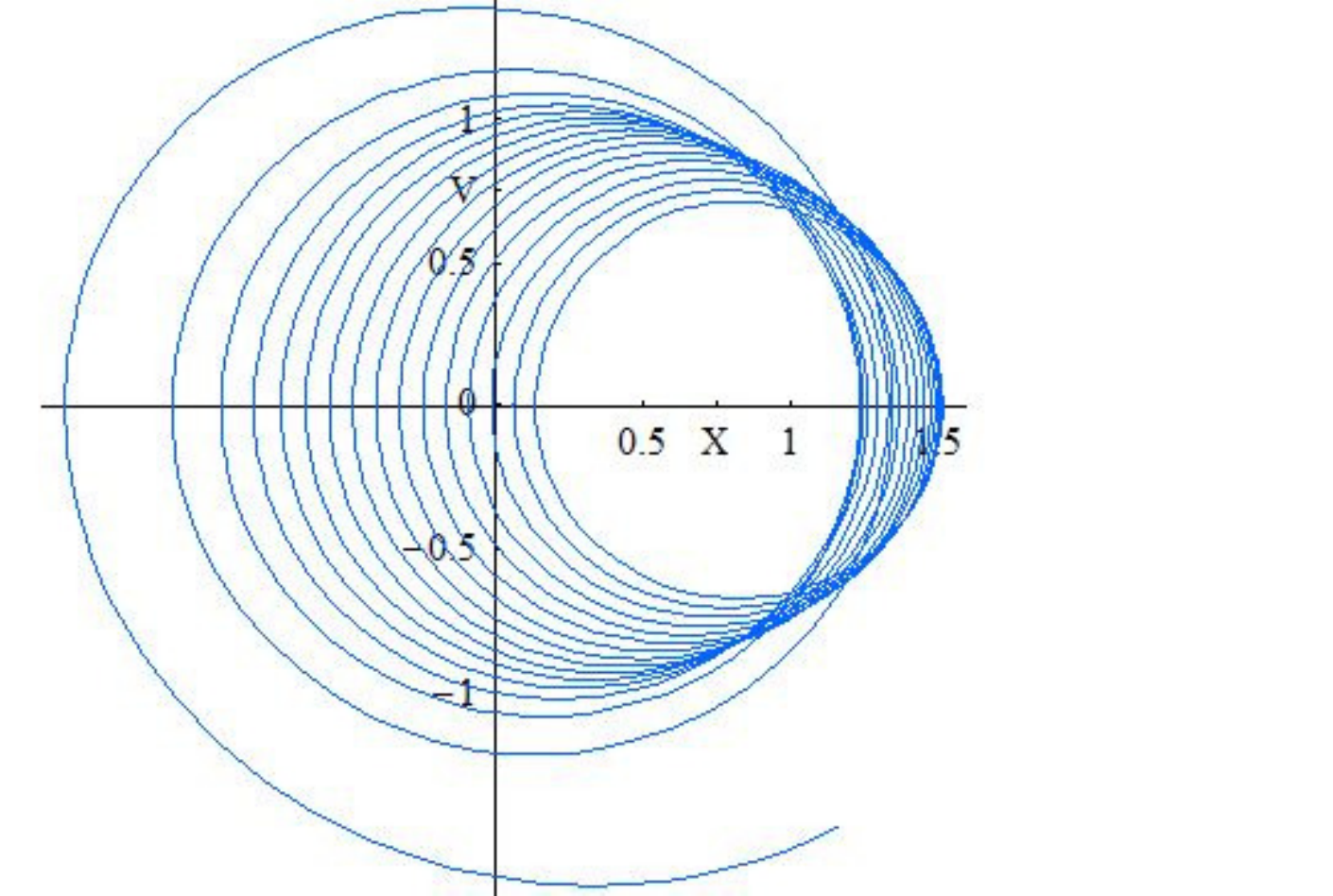}\includegraphics[width=3cm]{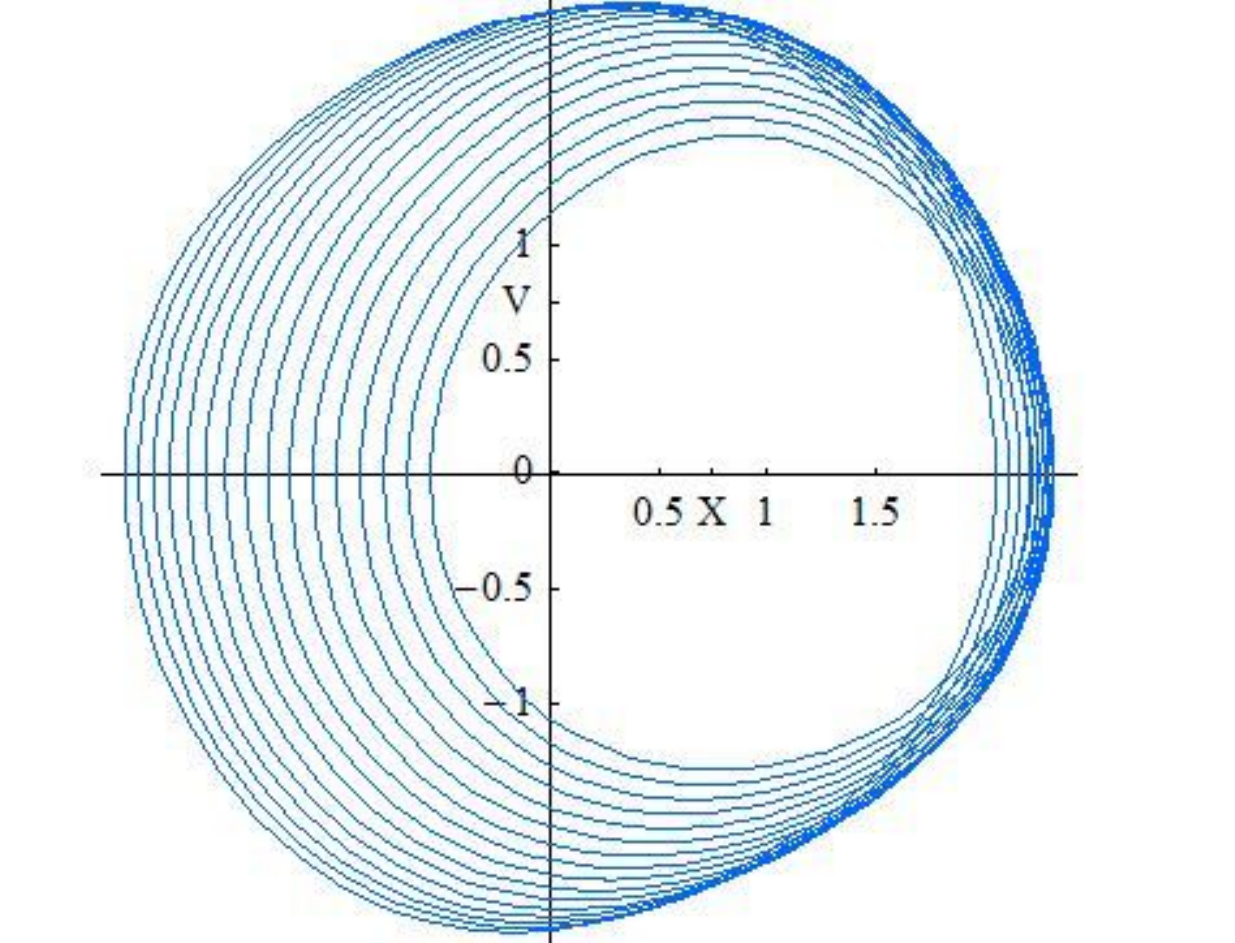}
  \includegraphics[width=3cm]{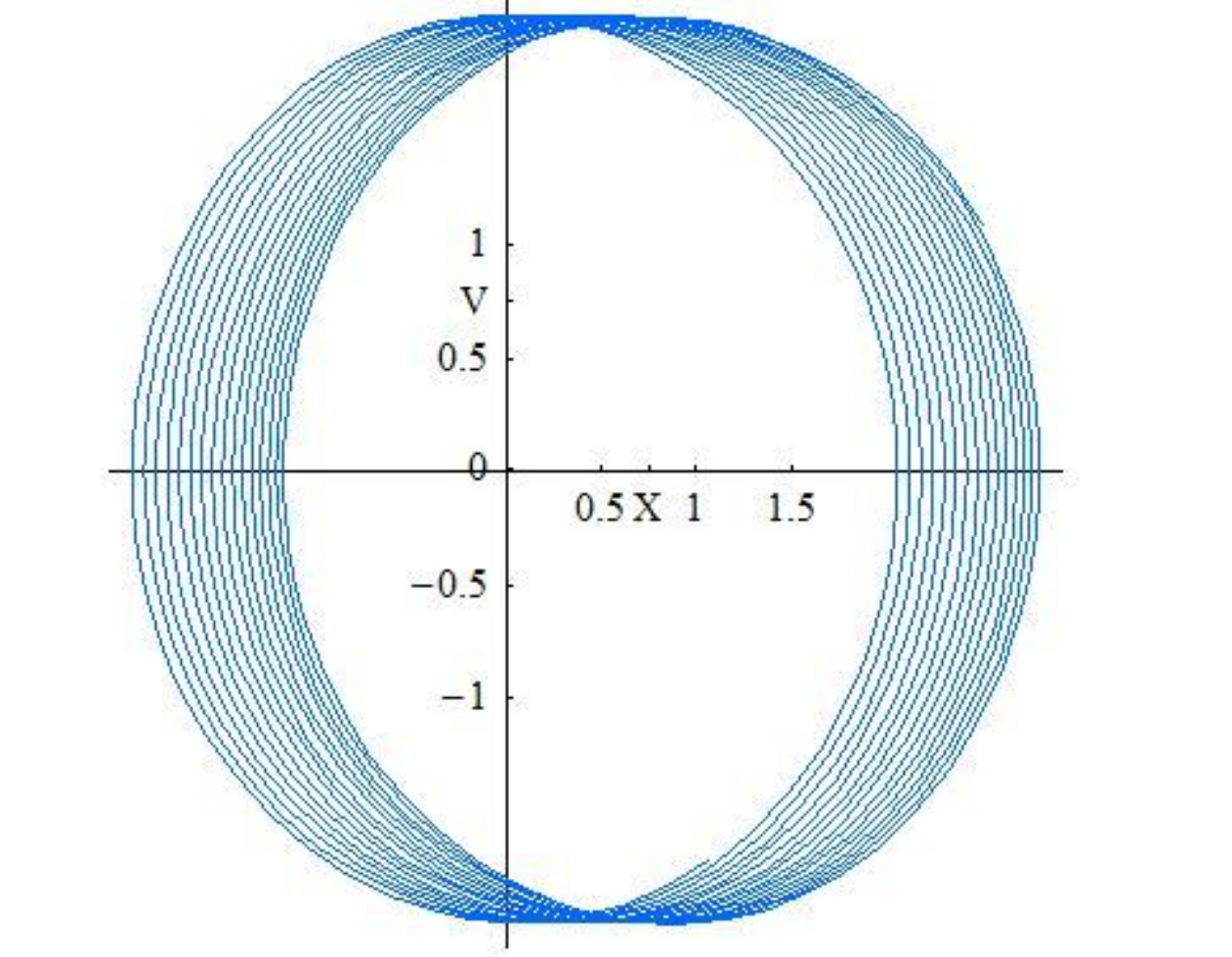}
  \end{center}
  \caption{ phase diagram :  forced Van der Pol generalized  oscillator, 
  forced  Van der Pol  oscillator and  forced Rayleigh  oscillator for $\varepsilon=10^{-1}, E=1 ,\Omega=10^{-2}.$ }
  \label{fig:2}
  \end{figure}

\begin{figure}[htbp]
 \begin{center}
 \includegraphics[width=3cm]{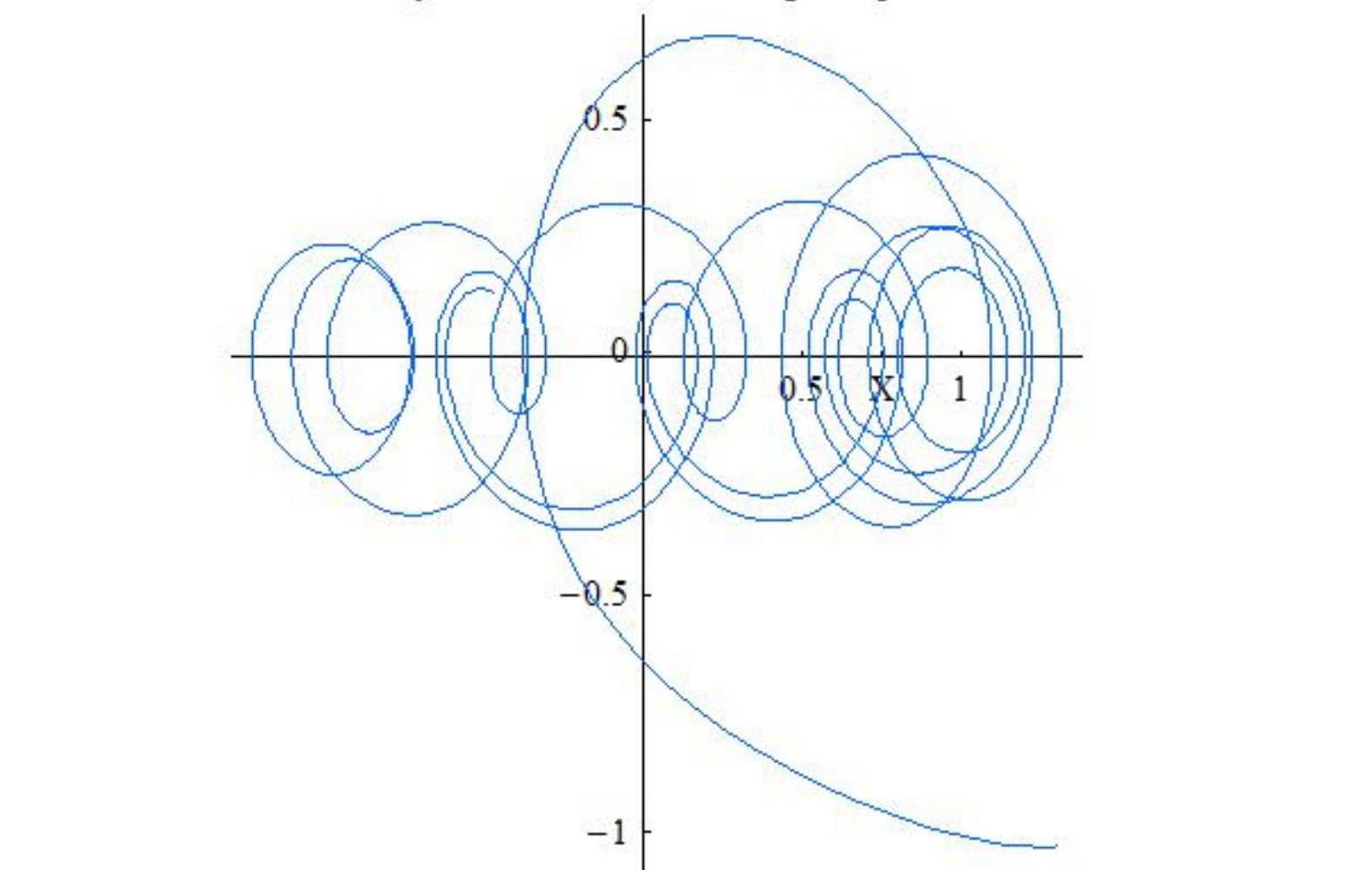}\includegraphics[width=3cm]{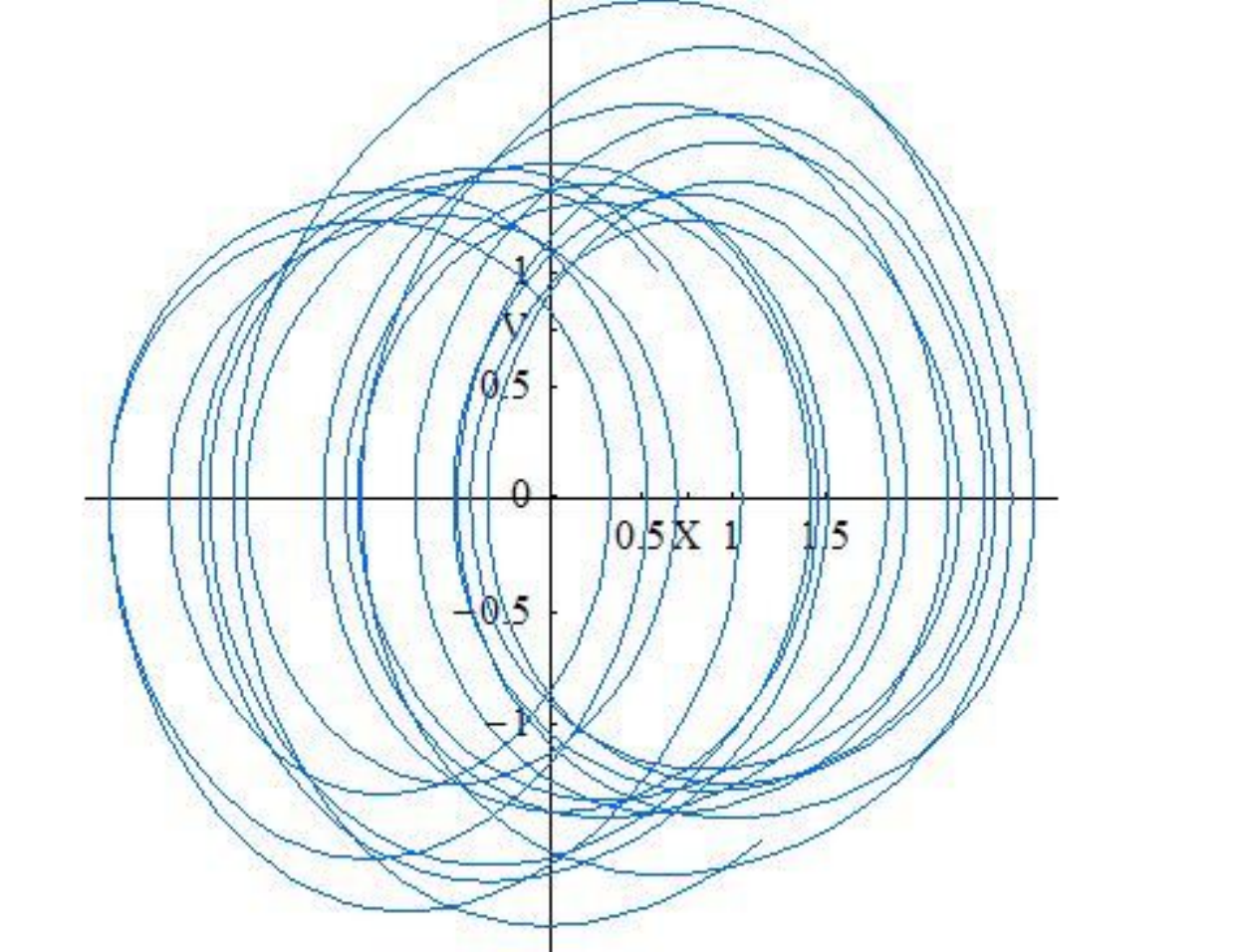}
 \includegraphics[width=3cm]{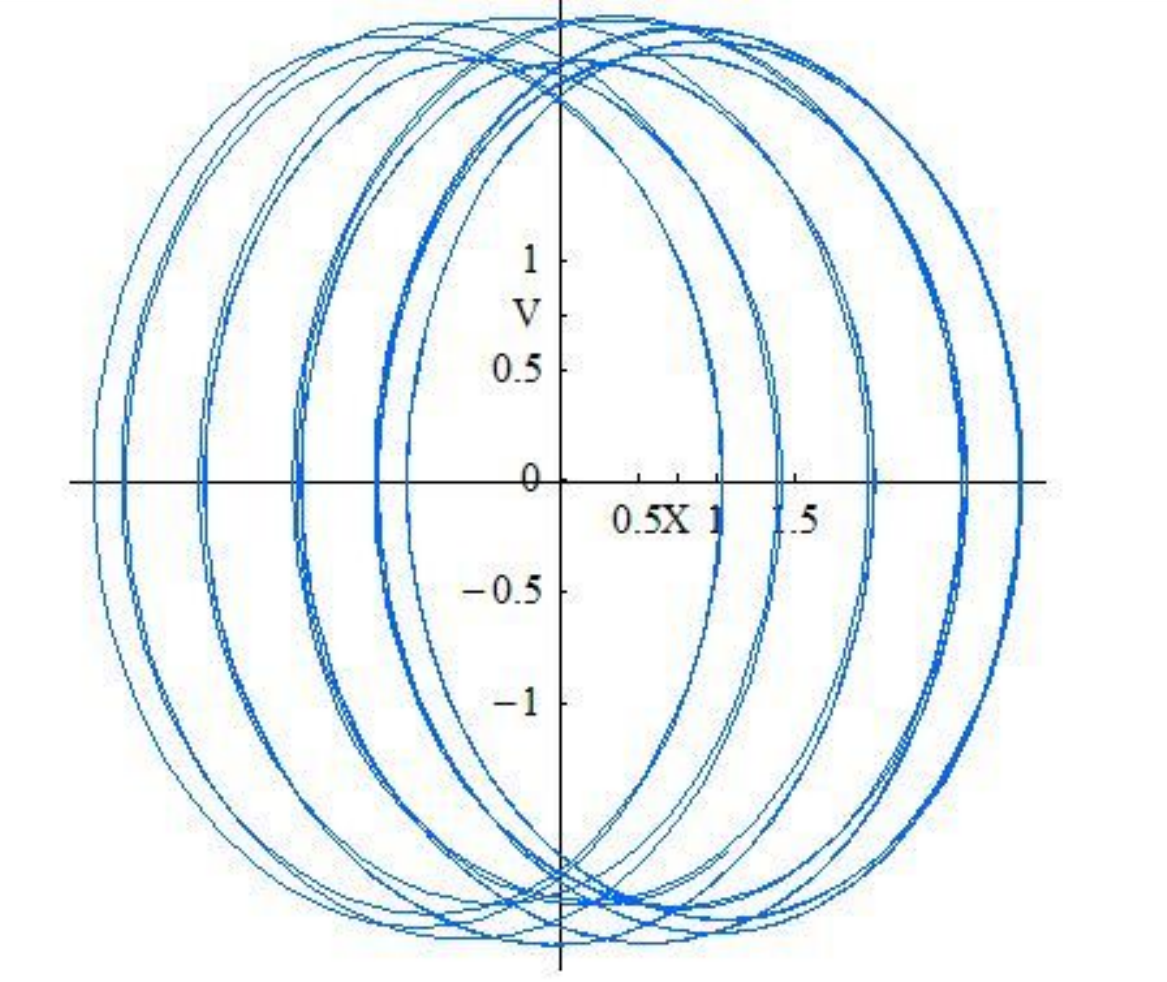}
\end{center}
 \caption{ phase  diagram :  forced Van der Pol generalized oscillator, 
  forced Van der Pol oscillator, and forced Rayleigh  oscillator for $\varepsilon=10^{-1}, E=1 ,\Omega=10^{-1}.$} 
 \label{fig:3}
 \end{figure}
 \newpage
For each one of the figures (\ref{fig:1}),   (\ref{fig:2}) and (\ref{fig:3}),
we have, on  the left, the  phase   diagram of Van der Pol generalized oscillator, 
in the middle, the phase  diagram  of Van der Pol oscillator   and on  the right, 
the phase diagram  of Rayleigh oscillator. These figures   show us, progressively 
when one  increases the magnitudeof $E$ and  $\Omega$,   the phase 
portrait goes from periodic condition,   almost periodic  
to chaotic condition. They   show  us also, the effects of 
the control parameters  $a$ and  $b$  on the system.

\begin{figure}[htbp]
 \begin{center}
 \includegraphics[width=5cm]{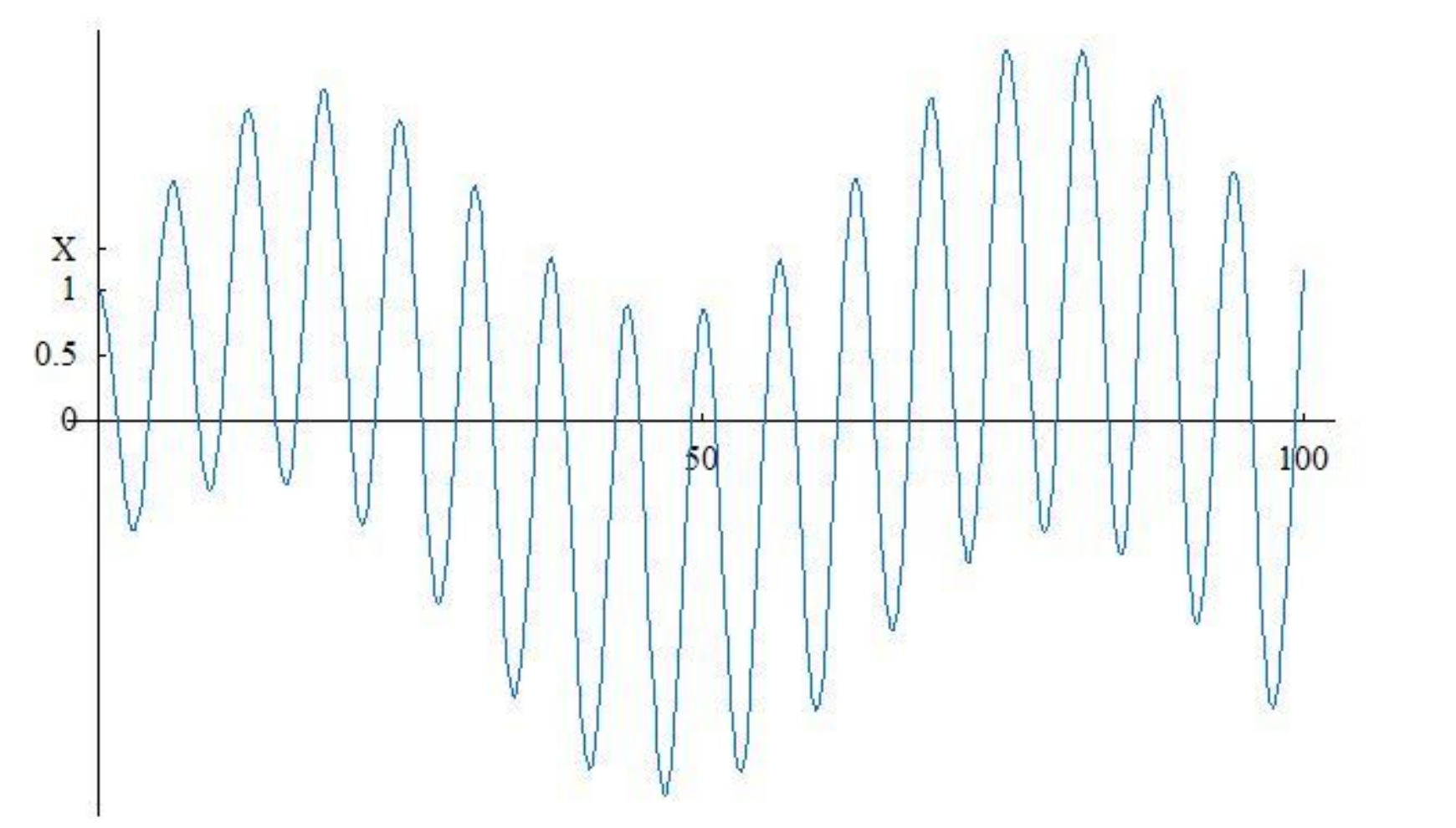}  \includegraphics[width=5cm]{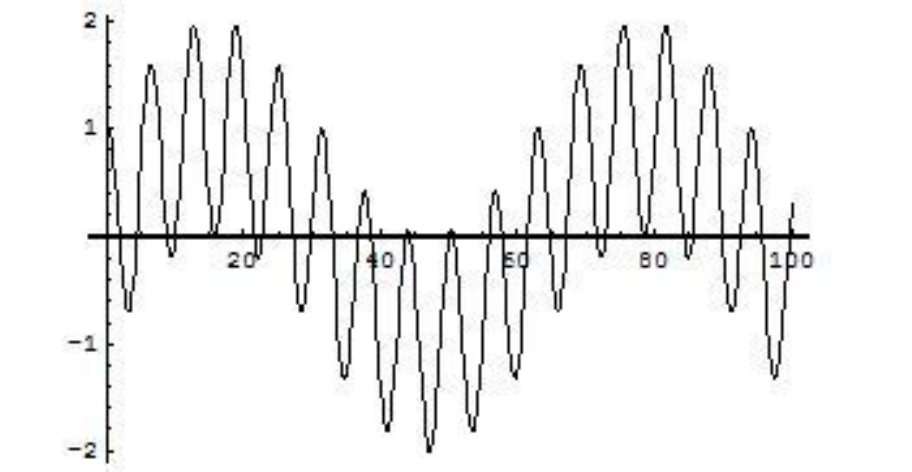}
 \end{center}
 \caption{ $x(t)$, for : $\varepsilon=10^{-1}$,  $\Omega=10^{-1}$, $E=1$, $a=1$, $b=1$.} 
 \label{fig:4}
 \end{figure}

\begin{figure}[htbp]
 \begin{center}
 \includegraphics[width=5cm]{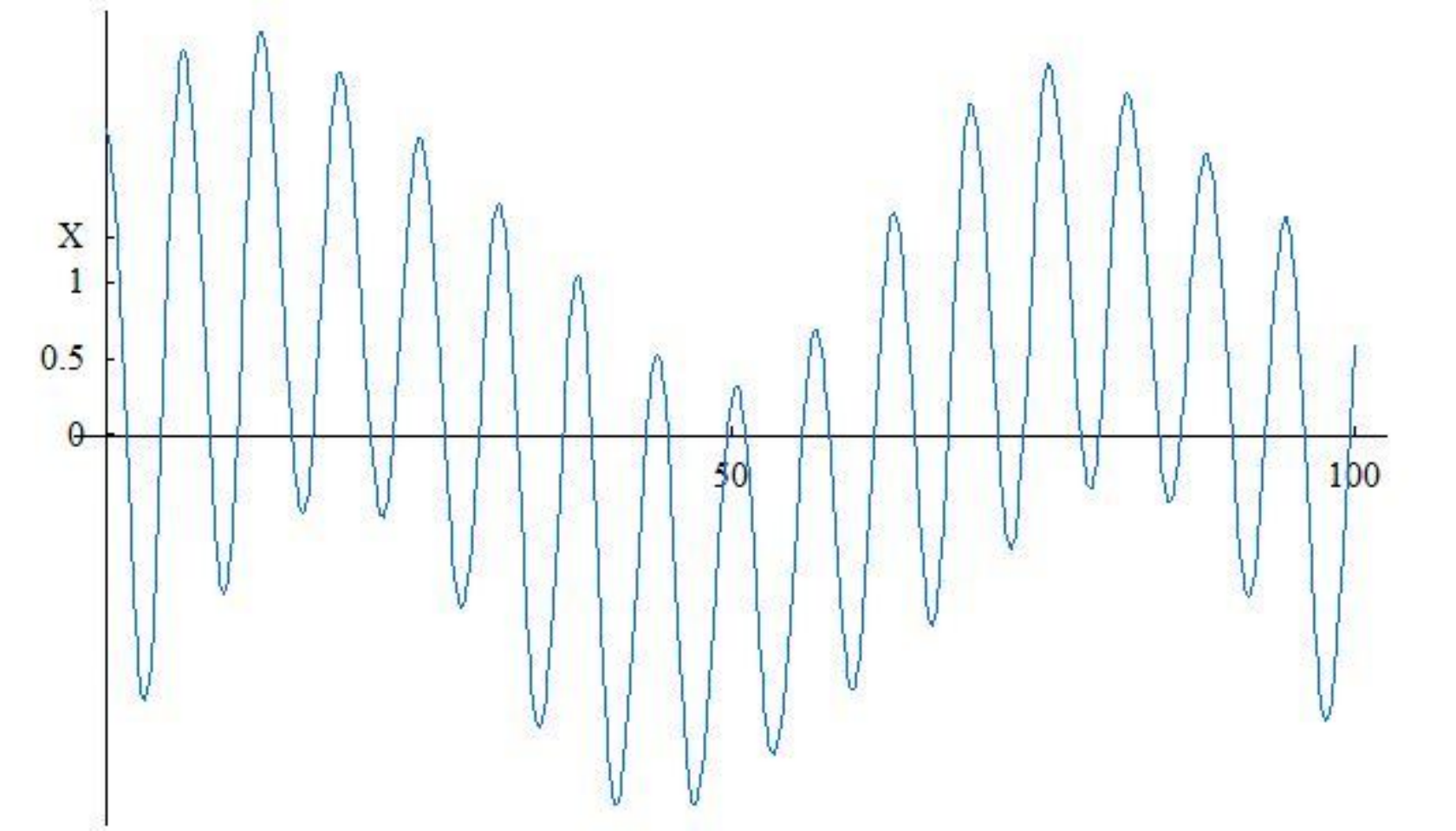}\includegraphics[width=5cm]{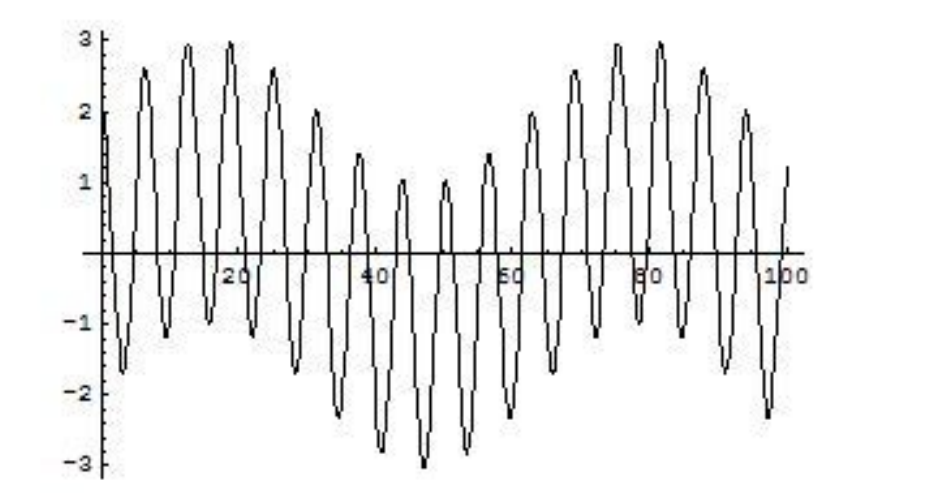}
 \end{center}
 \caption{ $x(t)$  for : $\varepsilon=10^{-1}$, $\Omega=10^{-1}$, $E=1$, $a=1$, $b=0$.}
 \label{fig:5}
 \end{figure}

\begin{figure}[htbp]
 \begin{center}
 \includegraphics[width=5cm]{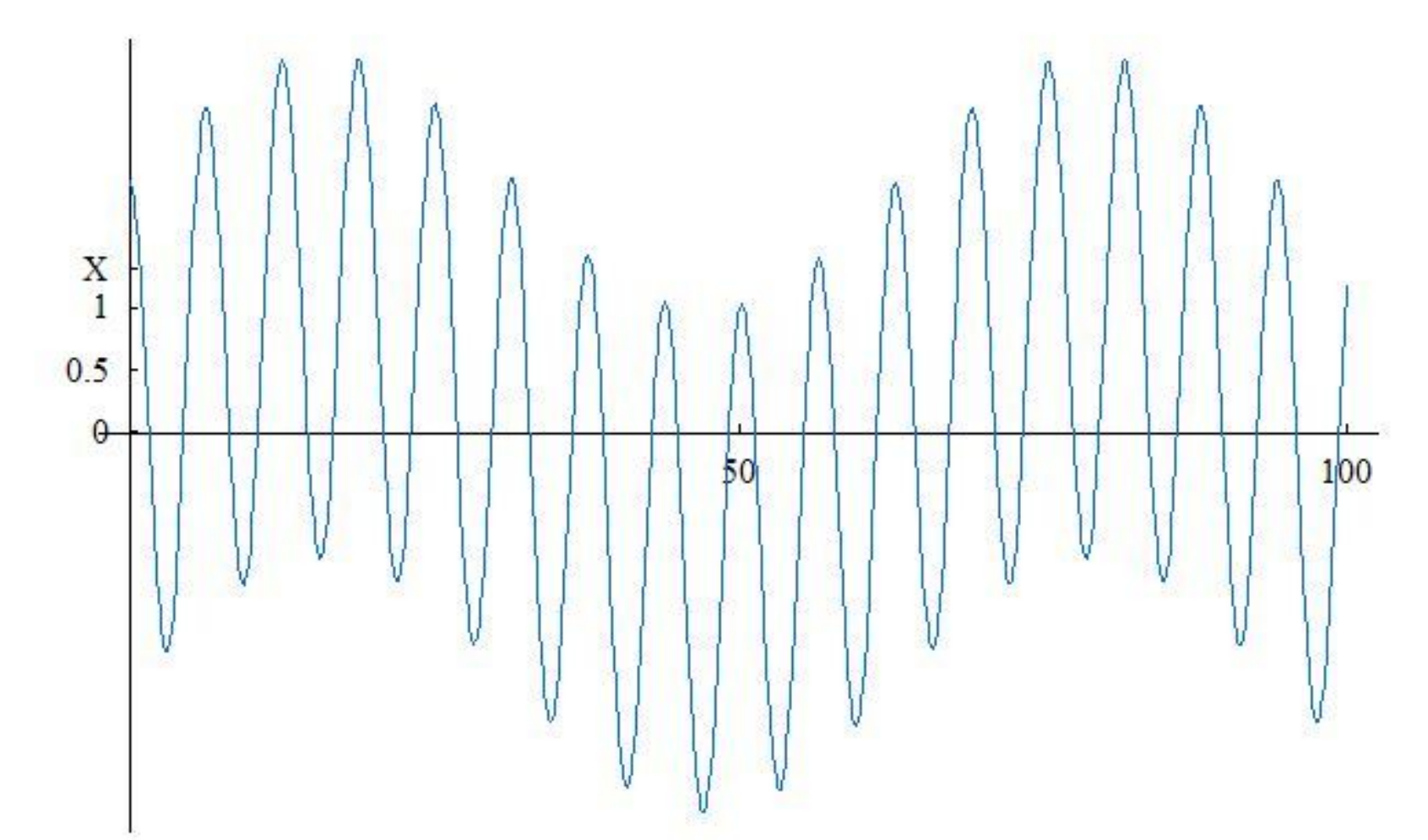}\includegraphics[width=5cm]{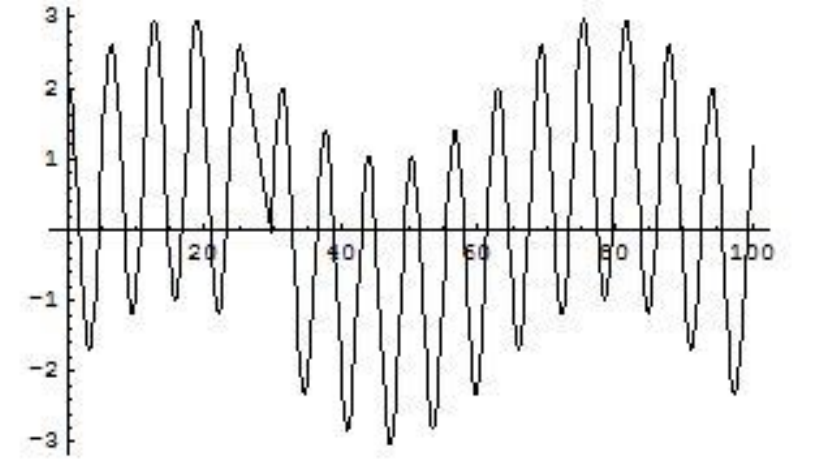}
 \end{center}
 \caption{ $x(t)$  for : $\varepsilon=10^{-1}$, $\Omega=10^{-1}$,  $E=1$, $a=0$, $b=\frac{1}{3}$.}
 \label{fig:6}
 \end{figure}

\begin{figure}[htbp]
 \begin{center}
 \includegraphics[width=5cm]{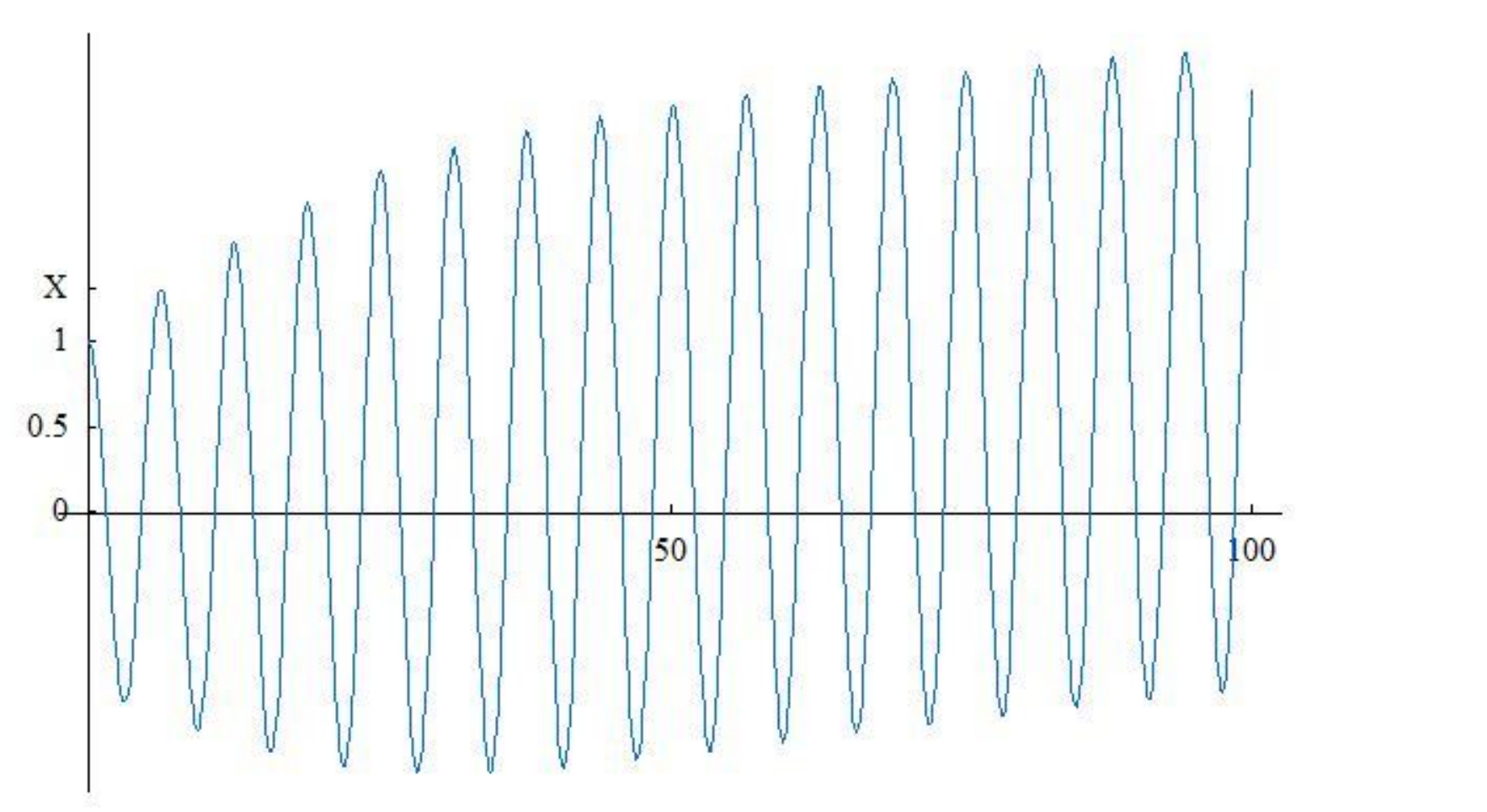}\includegraphics[width=5cm]{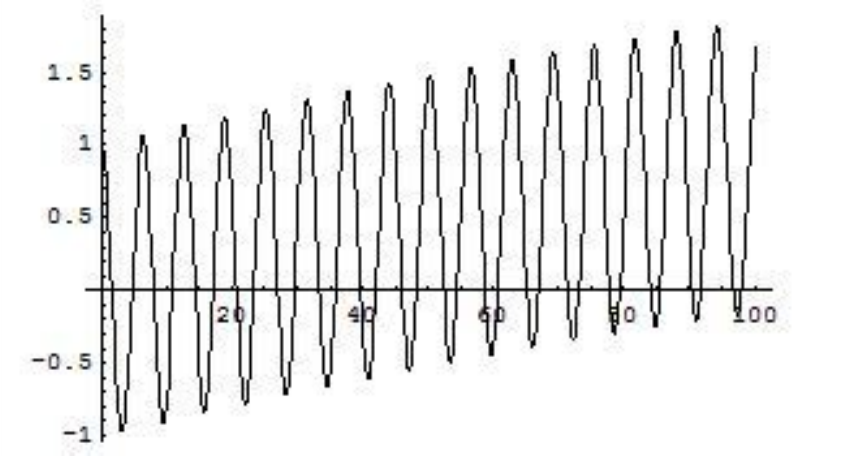}
 \end{center}
 \caption{ $x(t)$  for :  $\varepsilon=10^{-1}$, $\Omega=10^{-2}$,  $E=1$, $a=1$, $b=1$.}
 \label{fig:7}
 \end{figure}

\begin{figure}[htbp]
 \begin{center}
 \includegraphics[width=5cm]{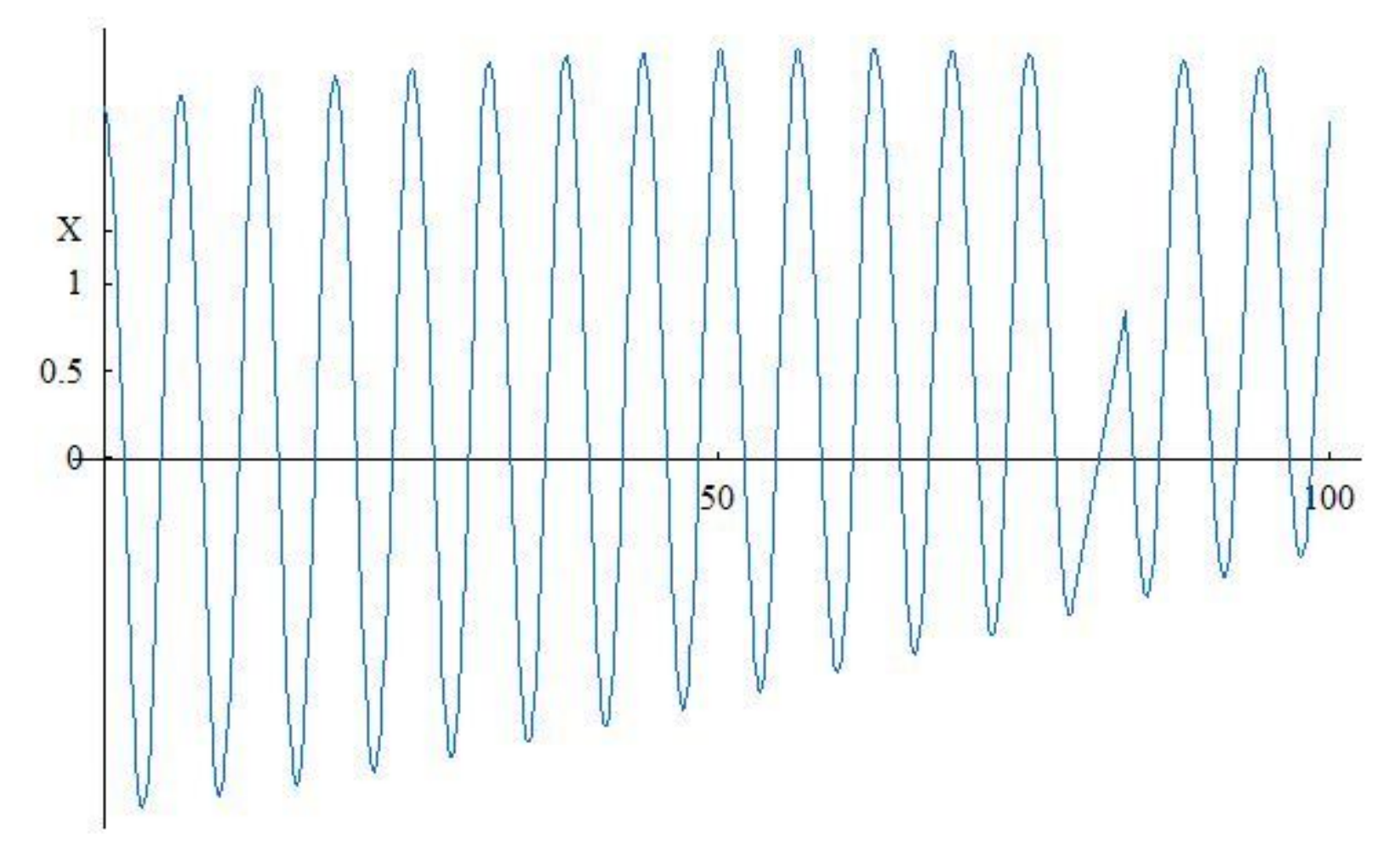} \includegraphics[width=5cm]{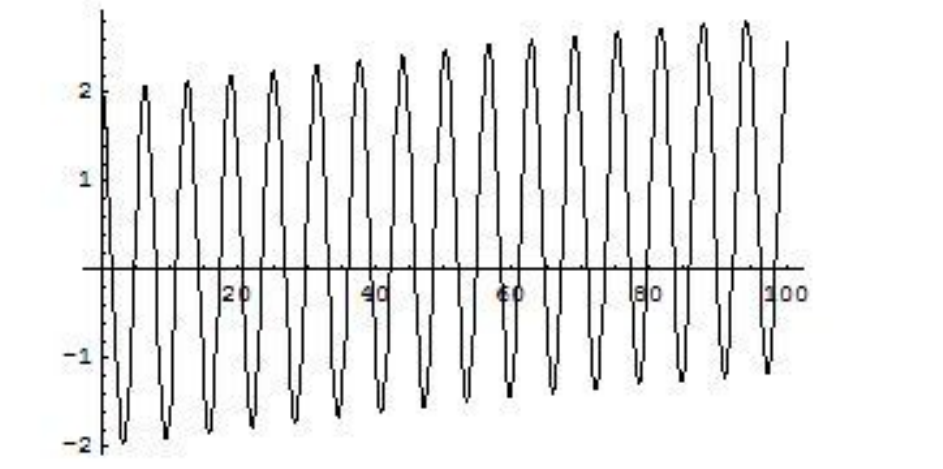}
 \end{center}
 \caption{ $x(t)$  for : $\varepsilon= 10^{-1}$, $ \Omega=10^{-2}$, $E=1$, $a=1$,  $b=0$.}
 \label{fig:8}
 \end{figure}

\begin{figure}[htbp]
 \begin{center}
 \includegraphics[width=5cm]{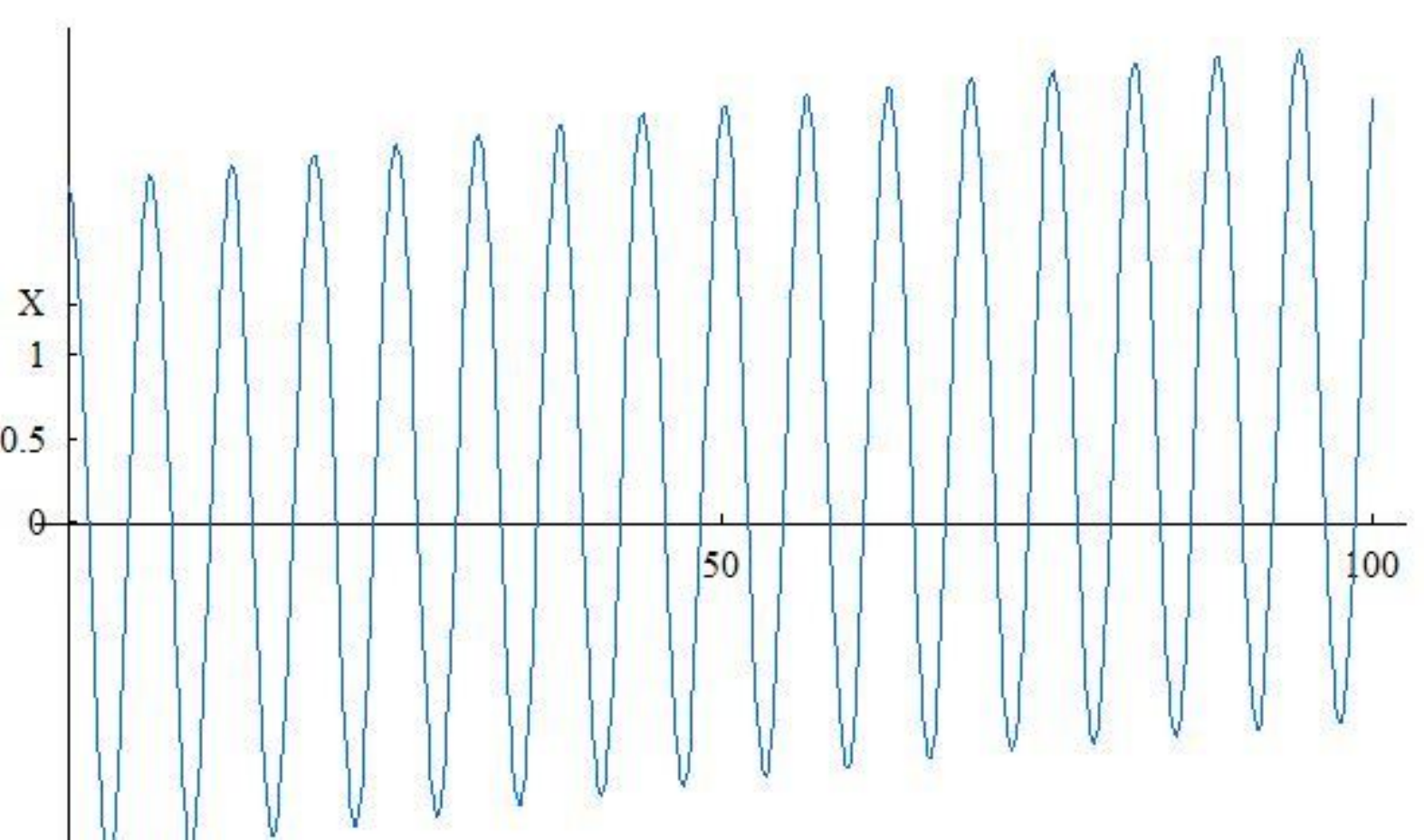}\includegraphics[width=5cm]{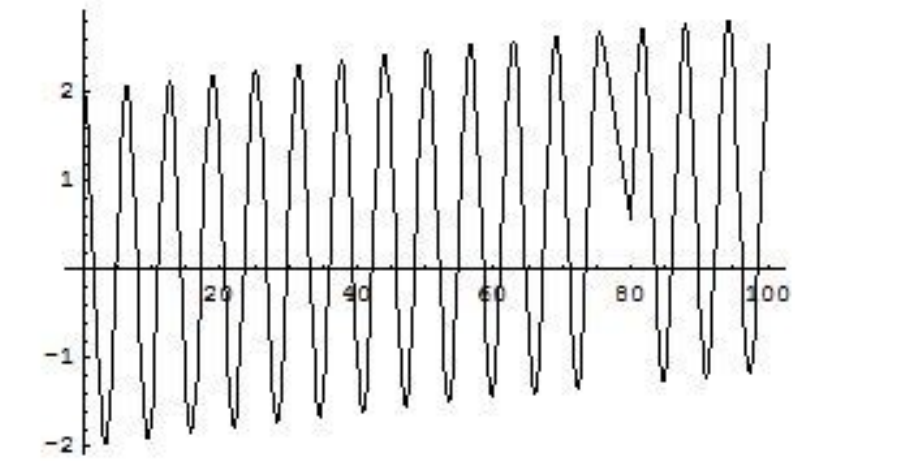}
 \end{center}
 \caption{ $x(t)$  for :$\varepsilon=10^{-1}$,  $\Omega=10^{-2}$, $E=1$, $a=0$, $b=\frac{1}{3}$.}
 \label{fig:9}
 \end{figure}

\begin{figure}[htbp]
 \begin{center}
 \includegraphics[width=5cm]{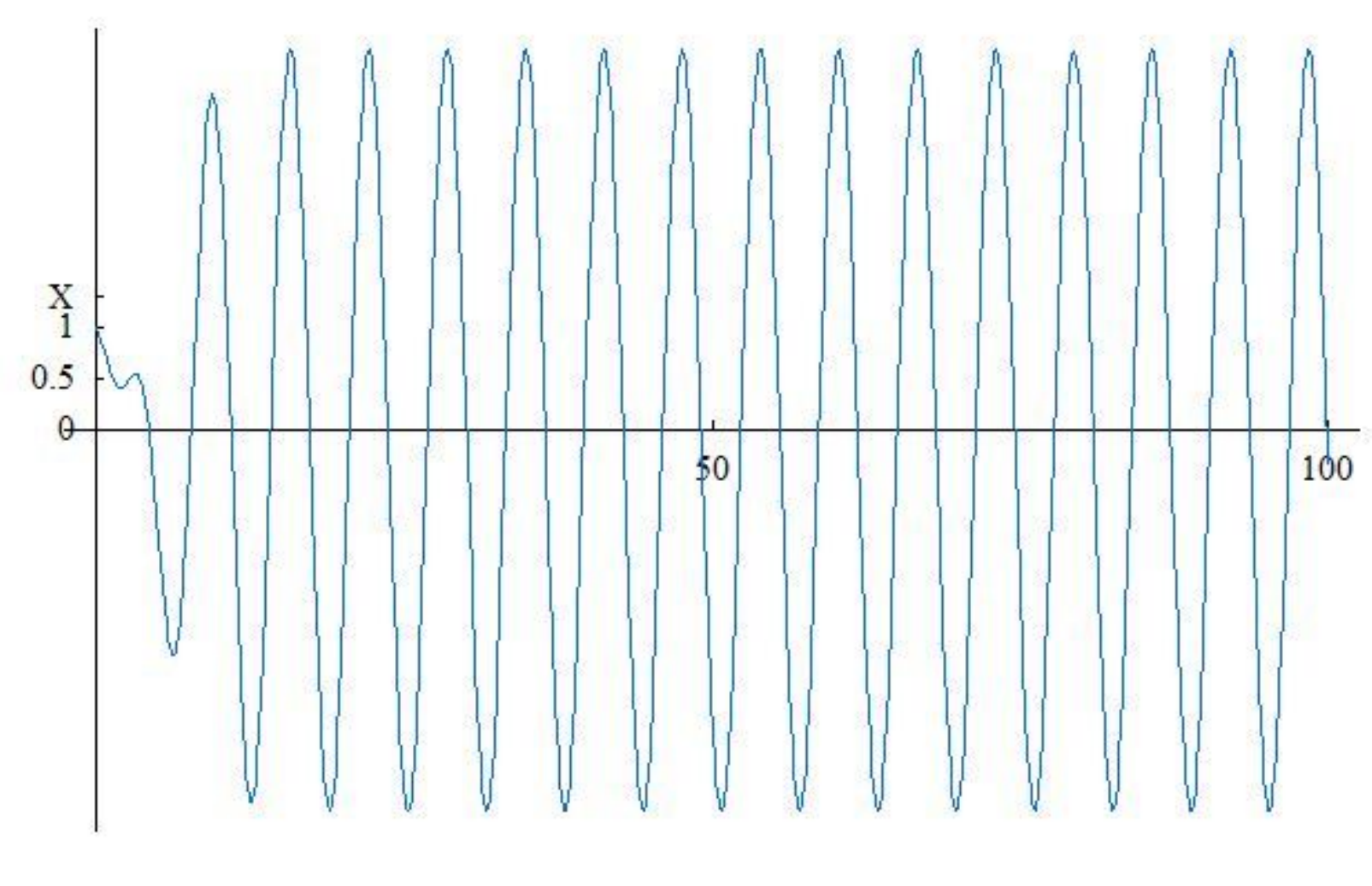}\includegraphics[width=5cm]{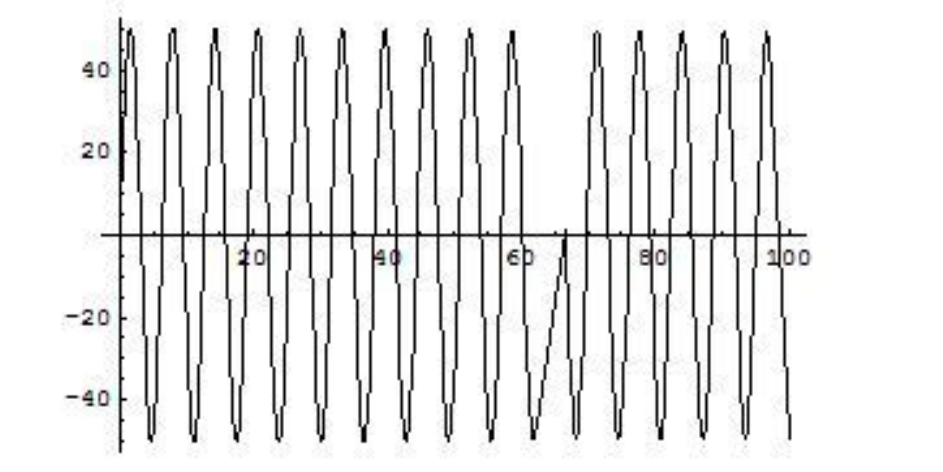}
 \end{center}
 \caption{ $x(t)$  for : $\varepsilon=10^{-1}$, $\Omega=99.10^{-2}$, $E=1$, $a=1$, $b=1$.}
 \label{fig:10}
 \end{figure}

\begin{figure}[htbp]
 \begin{center}
 \includegraphics[width=5cm]{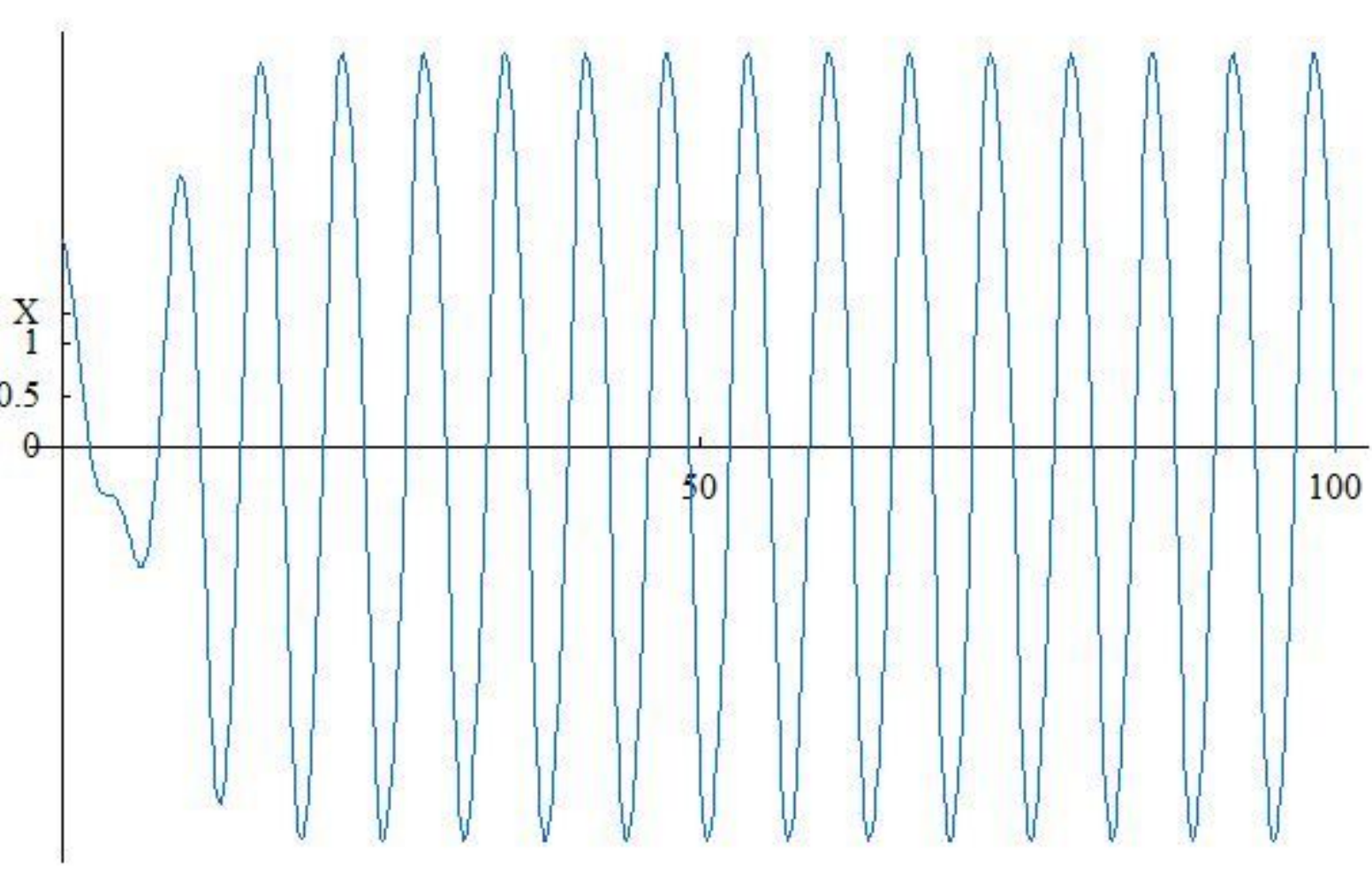}\includegraphics[width=5cm]{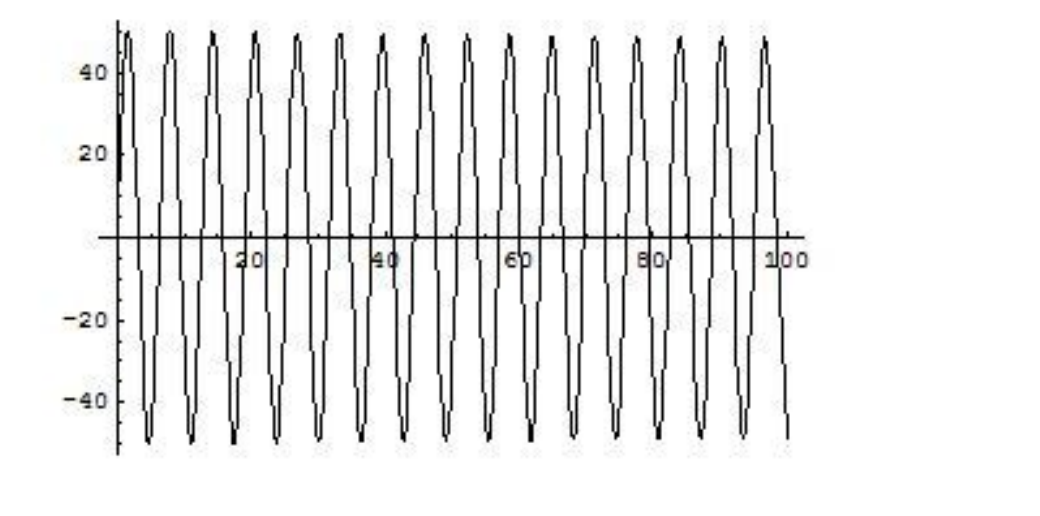}
 \end{center}
 \caption{ $x(t)$  for : $\varepsilon=10^{-1}$, $\Omega=99.10^{-2}$ , $E=1$, $a=1$, $b=0$.}
 \label{fig:11}
 \end{figure}

\begin{figure}[htbp]
 \begin{center}
 \includegraphics[width=5cm]{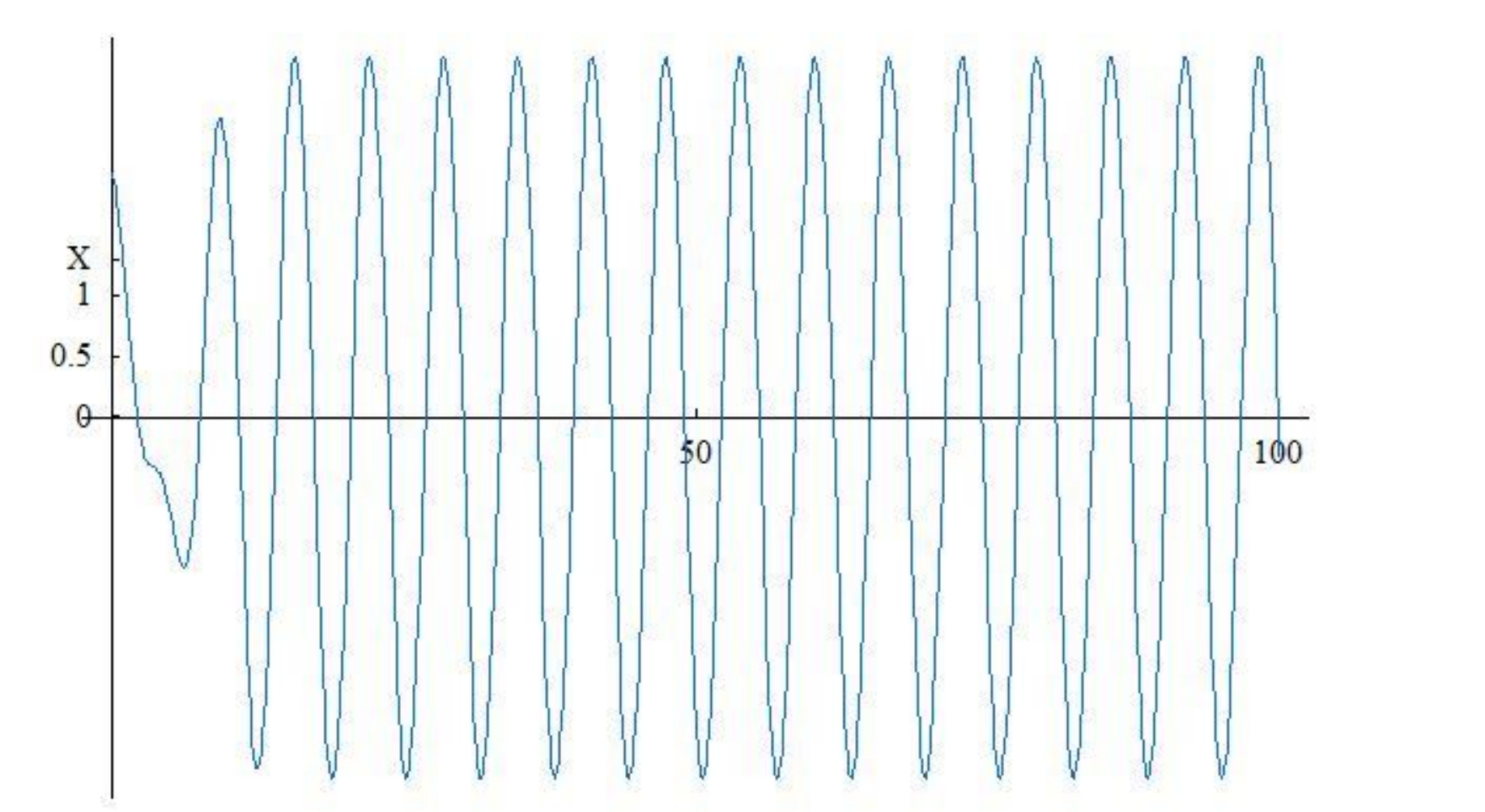}\includegraphics[width=5cm]{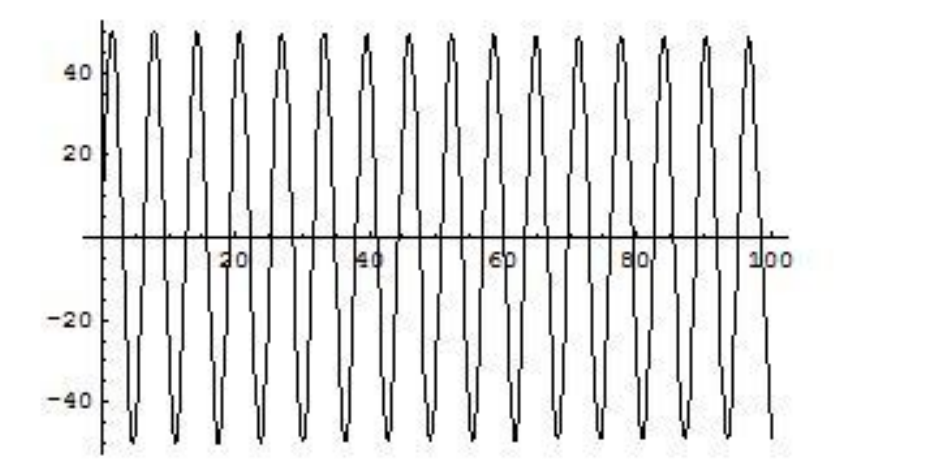}
 \end{center}
 \caption{ $x(t)$  for $\varepsilon=10^{-1},\Omega=99.10^{-2}, E=1, a=0, b=\frac{1}{3}$.}
 \label{fig:12}
 \end{figure}

\begin{figure}[htbp]
 \begin{center}
 \includegraphics[width=5cm]{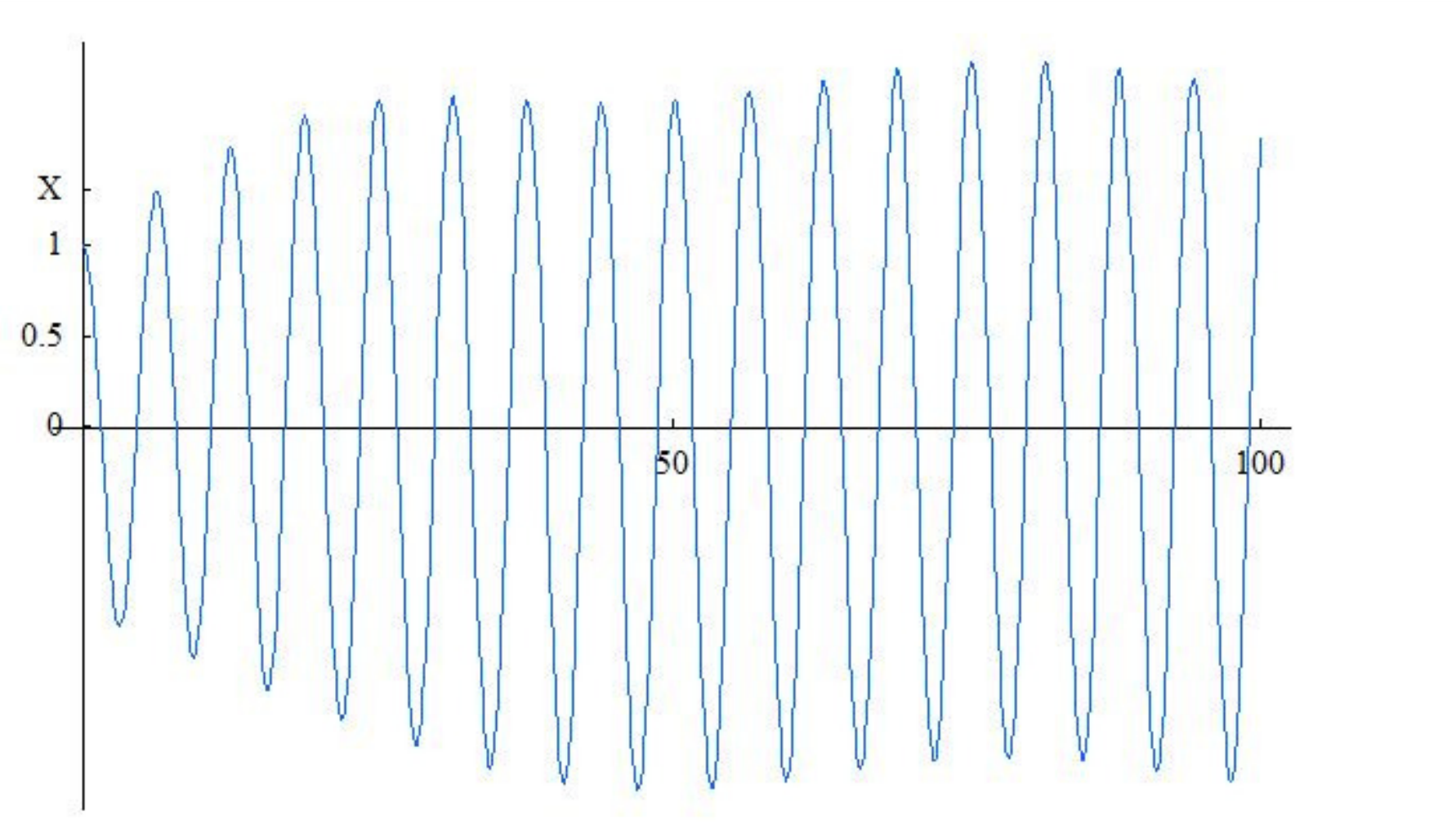}\includegraphics[width=5cm]{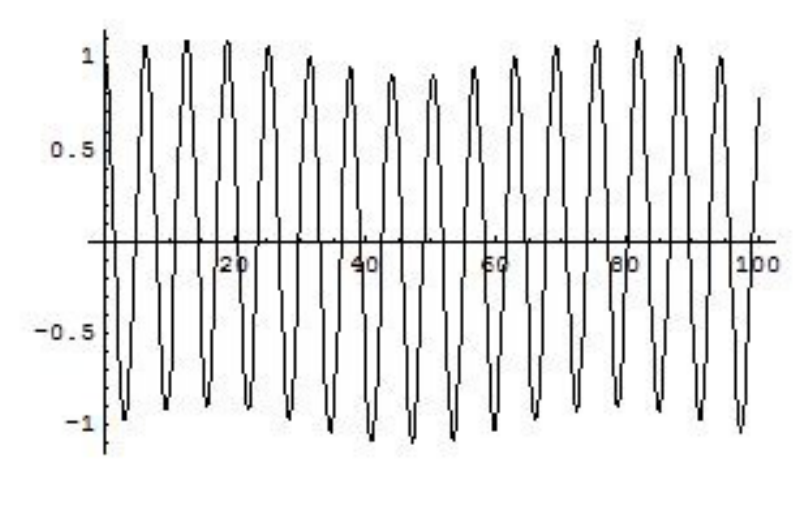}
 \end{center}
 \caption{ $x(t)$  for : $\varepsilon=10^{-1}$, $\Omega=10^{-1}$, $E=10^ {-1}$, $a=1$, $b=1$.}
 \label{fig:13}
 \end{figure}

\begin{figure}[htbp]
 \begin{center}
 \includegraphics[width=5cm]{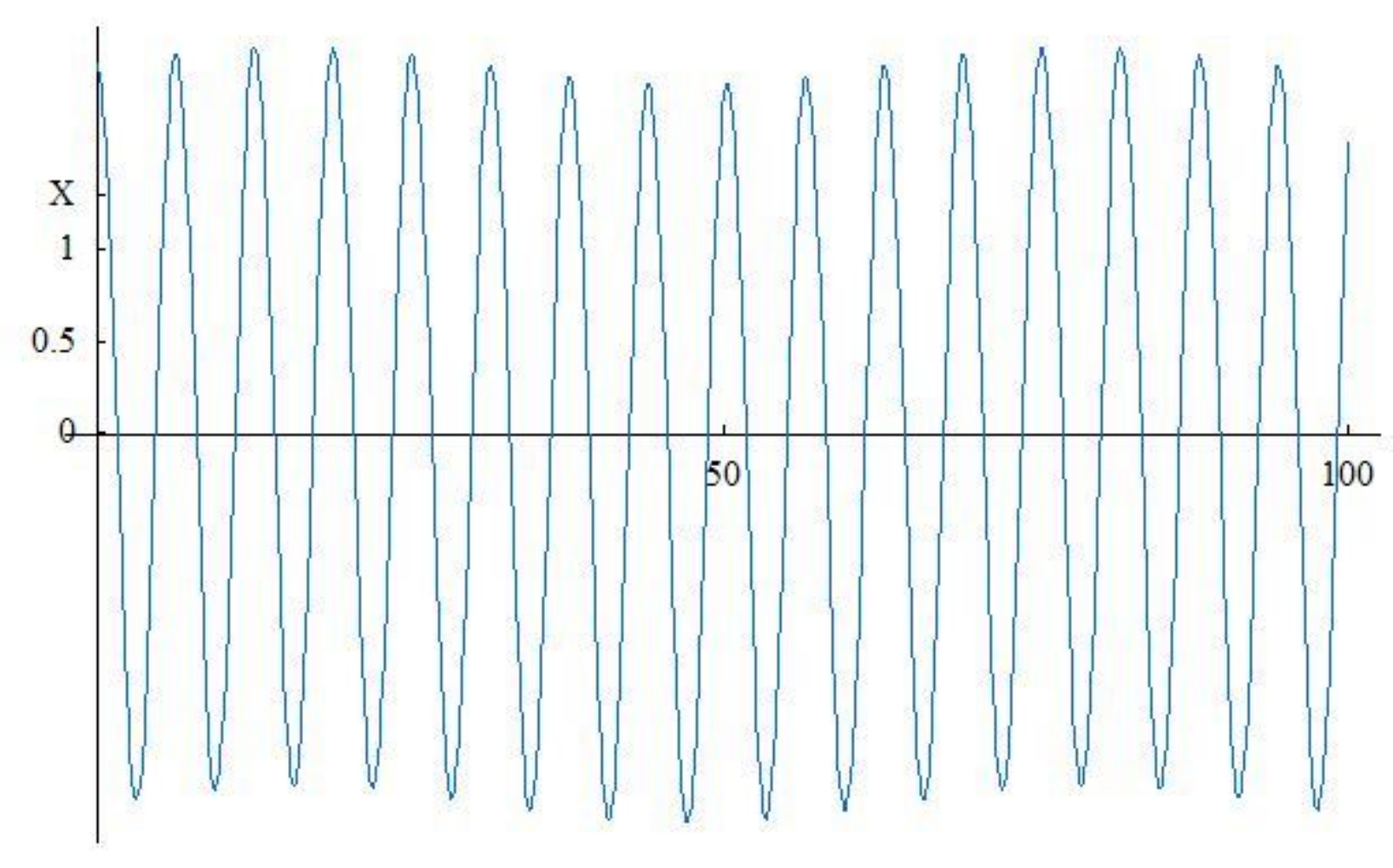}\includegraphics[width=5cm]{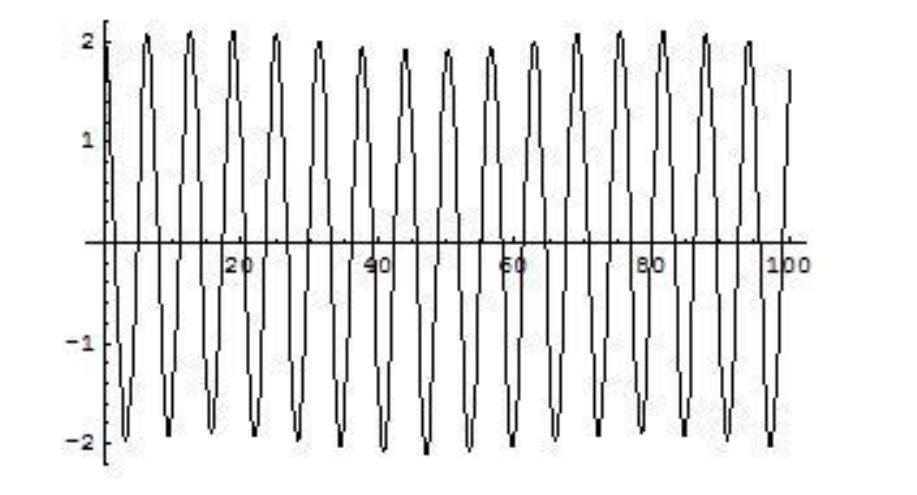}
 \end{center}
 \caption{ $x(t)$  for : $\varepsilon=10^{-1}$, $\Omega=10^{-1}$ , $E=10^ {-1}$, $a=1$,  $b=0$.}
 \label{fig:14}
 \end{figure}

\begin{figure}[htbp]
 \begin{center}
 \includegraphics[width=5cm]{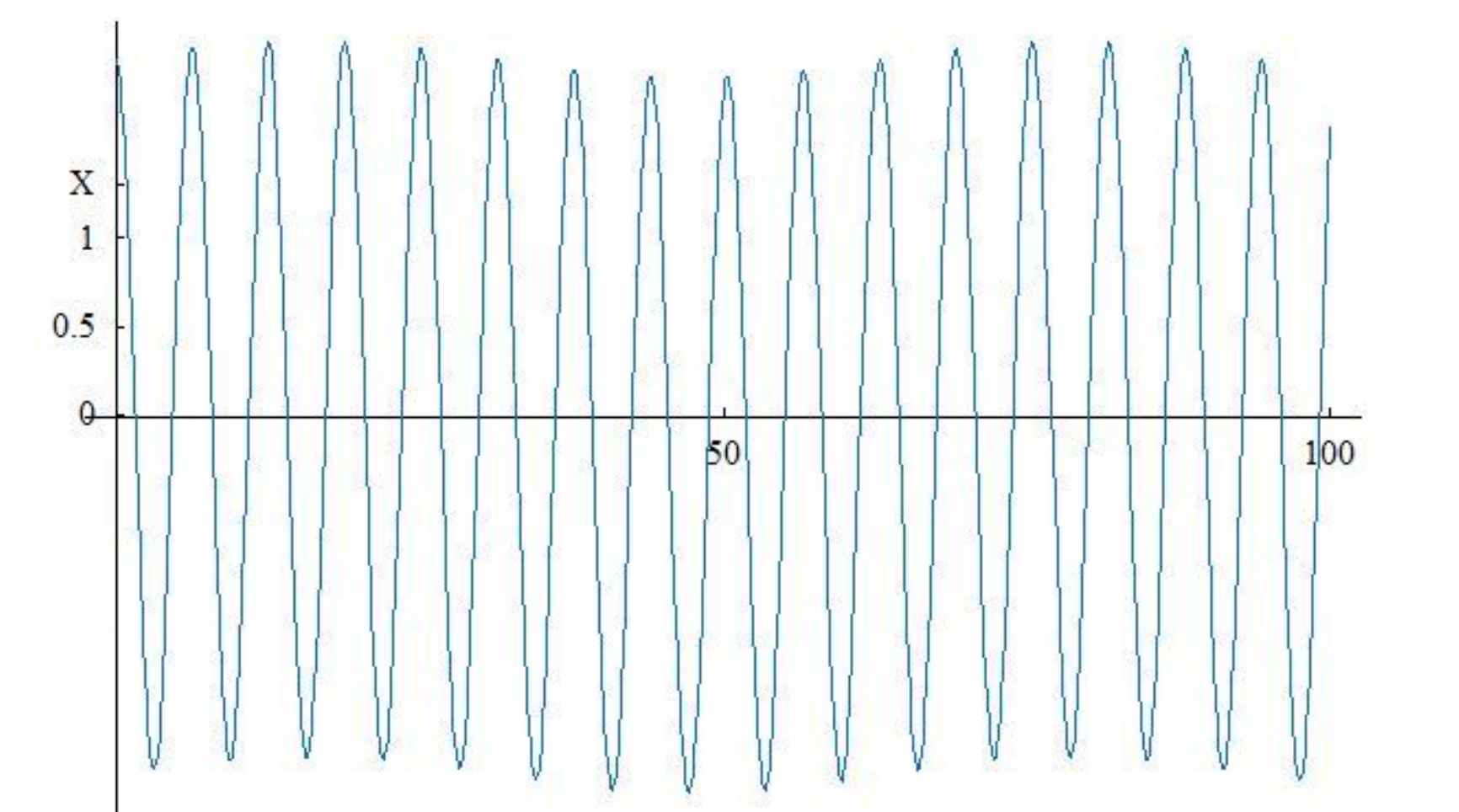}\includegraphics[width=5cm]{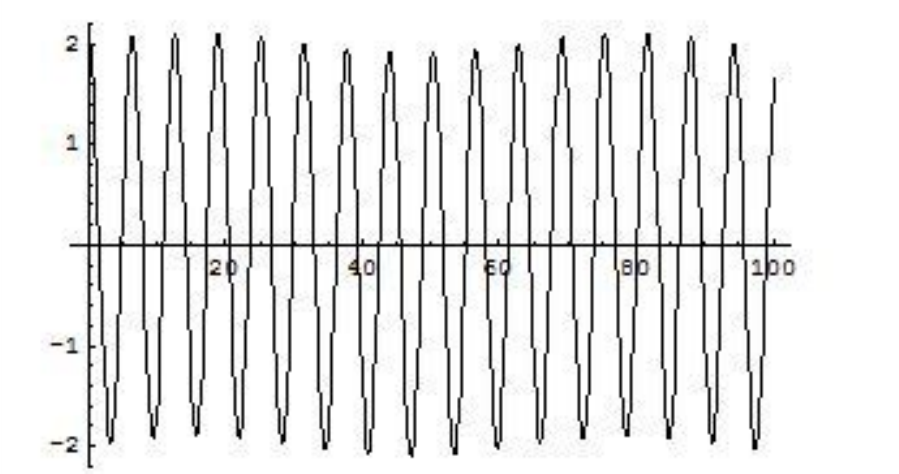}
 \end{center}
 \caption{ $x(t)$  for :$\varepsilon=10^{-1}$, $\Omega=10^{-1}$ , $E=10^ {-1}$, $a=0$, $b=\frac{1}{3}$.}
 \label{fig:15}
 \end{figure}

  \begin{figure}[htbp]
   \begin{center}
   \includegraphics[width=5cm]{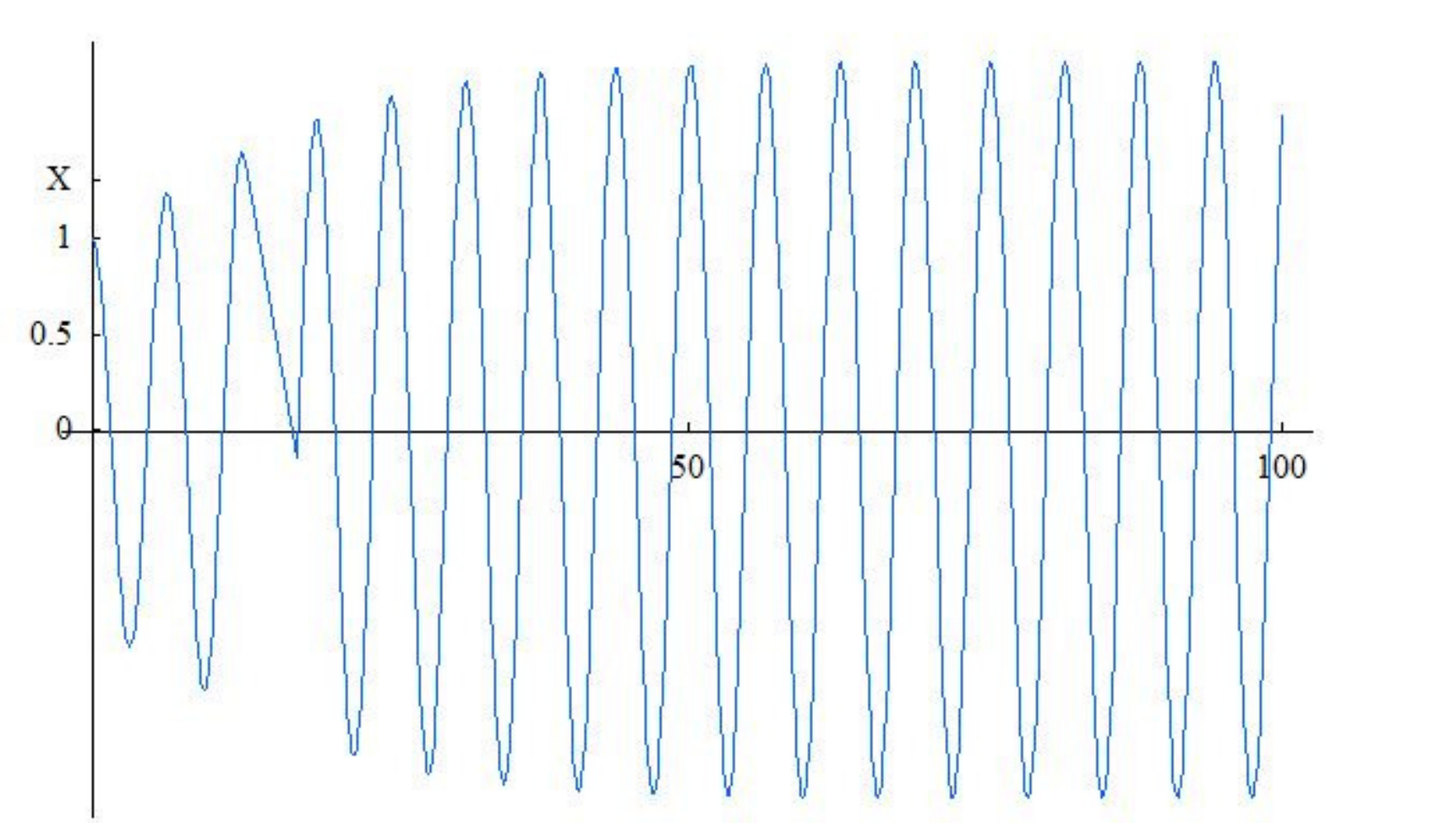}\includegraphics[width=5cm]{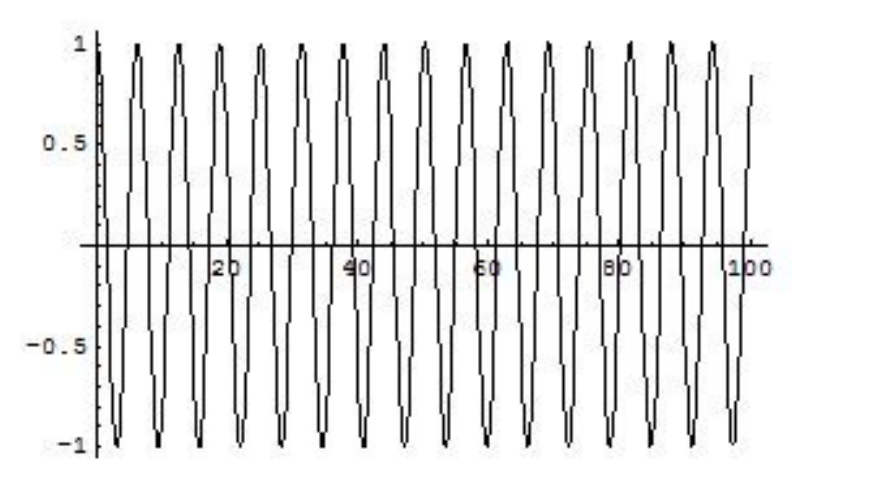}
   \end{center}
   \caption{ $x(t)$  for : $\varepsilon=10^{-1}$, $\Omega=10^{-2}$ , $E=10^ {-2}$, $a=1$, $b=1$.}
   \label{fig:16}
   \end{figure}

  \begin{figure}[htbp]
   \begin{center}
   \includegraphics[width=5cm]{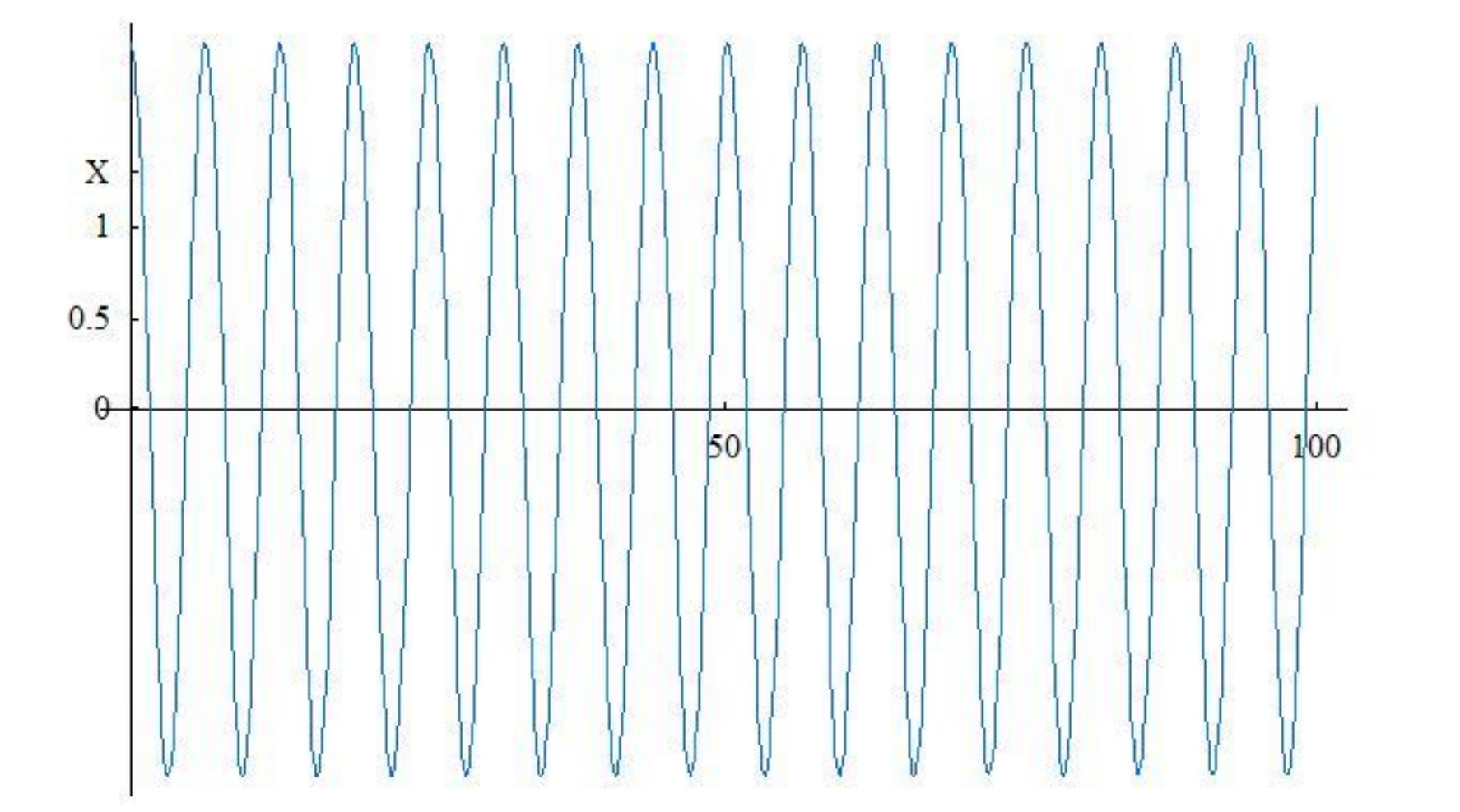}\includegraphics[width=5cm]{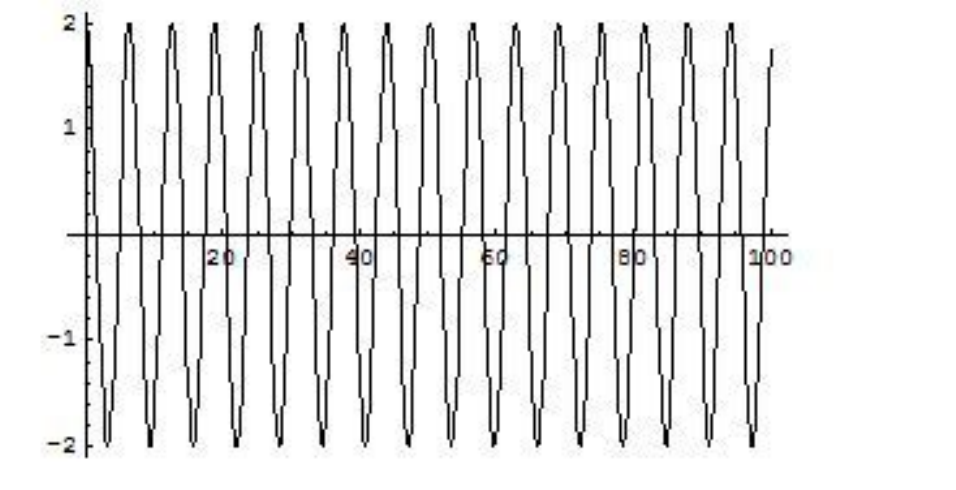}
   \end{center}
   \caption{ $x(t)$  for : $\varepsilon=10^{-1}$, $\Omega=10^{-2}$, $E=10^ {-2}$, $a=1$, $b=0$.}
   \label{fig:17}
   \end{figure}

 \begin{figure}[htbp]
  \begin{center}
  \includegraphics[width=5cm]{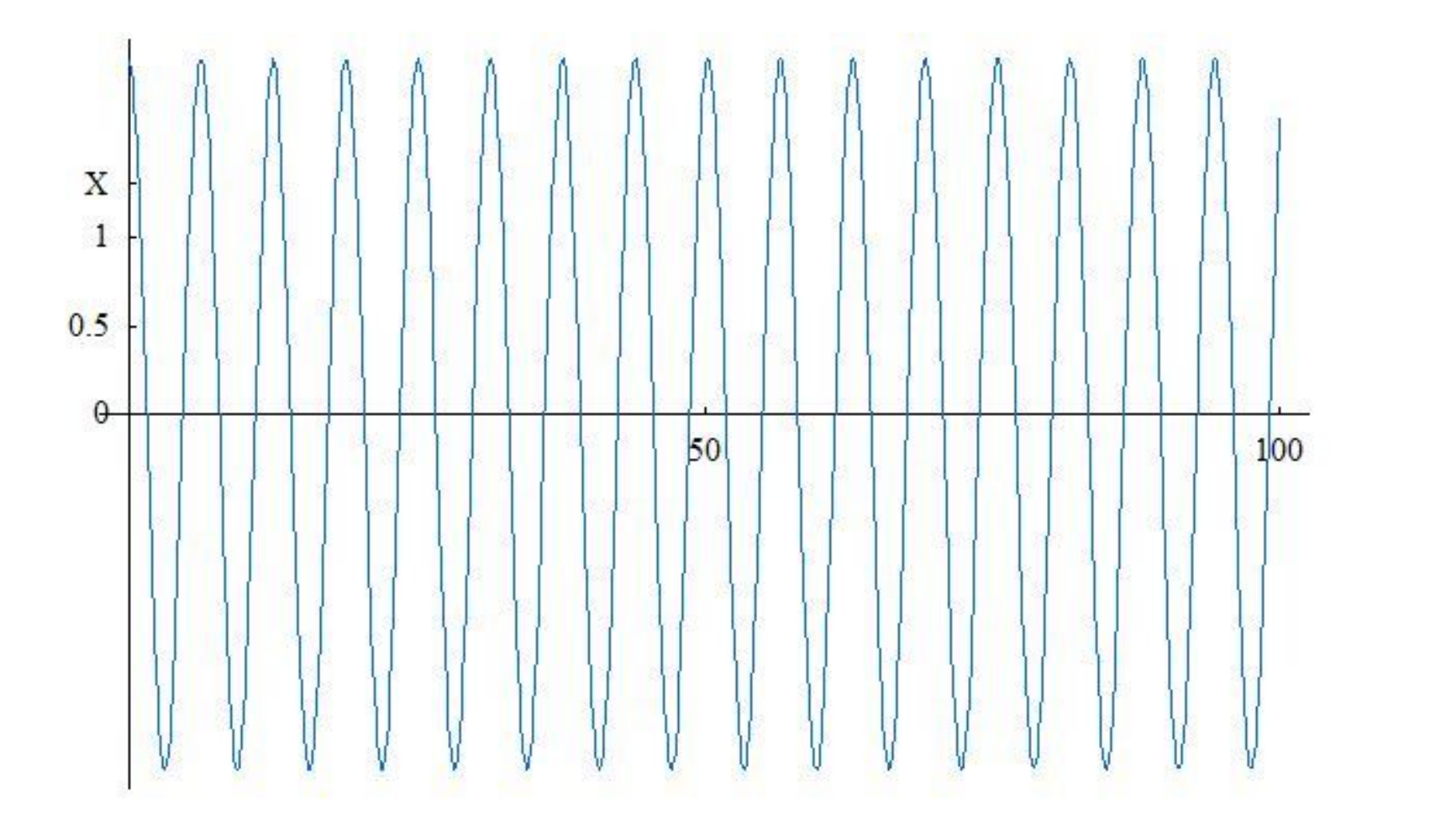}\includegraphics[width=5cm]{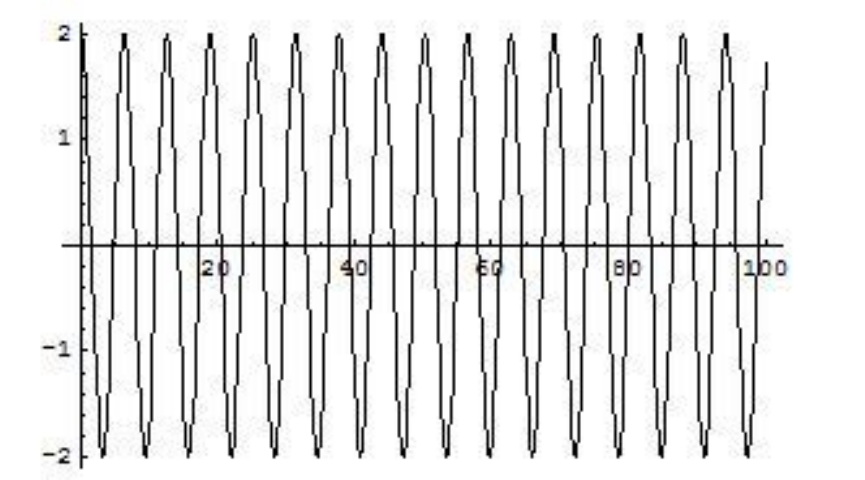}
  \end{center}
  \caption{ $x(t)$  for : $\varepsilon=10^{-1}$, $\Omega=10^{-2}$, $E=10^ {-2}$, $a=0$, $b=\frac{1}{3}$.}
  \label{fig:18}
  \end{figure}

\newpage
As for the figures  from ( \ref{fig:4})  to  (\ref{fig:18}) we have the 
graph of the exact solution of Equation  (\ref{eq27})  on the left 
and the graph of the approximate solution Equation (\ref{eq39})   on the right. 
The chaotic condition noticed in Figure (\ref{fig:3}) is confirmed in real 
space through the  behavior of the curves  in the Figures 
(\ref{fig:4}), (\ref{fig:5}) and (\ref{fig:6})respectively
(Forced Van der Pol generalized, Forced  Van der Pol and Forced Rayleigh oscillators).
Furthermore the  quasi-periodic  oscillation is noticed in the behavior of the curves of the 
Figures (\ref{fig:7}), (\ref{fig:8}) and (\ref{fig:9}).

The second term $\left(\frac{E\sin{\Omega t}}{1-\Omega^{2}}\right)$  of the solution Equation
(\ref{eq39}) 
show the  appearance of the resonance for $ \Omega\simeq1$.
This behavior is illustrated by the figures \ref{fig:10}), \ref{fig:11}) and  \ref{fig:12})
where the  dynamic system's amplitude  is increasing.
We see through each figure that the approximate solution found  approaches
more or less  the exact  solution, which justifies that ours   result are 
optimal.

The equations of system (\ref{eq40})  show us that the phase initial of dynamic system  is a 
constant
and the amplitude  $r$ is  function of both the  time and the control parameters of system.
We chose this initial condition equal to zero to simplify our simulation. Also, the first equation
 of the equations system (\ref{eq40}) above, gives the stable cycle limit  radius that is only 
 a function of the parameters $a$  and $b$. They show also that there is the 
 occurrence of the Hopf's classical bifurcation.

\section{Conclusion}
We recalled the outline of the method of the renormalization group method for ordinary differential equations
$(ODEs)$ which provides in addition to the solution, the renormalization group $(EGR)$ which leads to the
determination of the amplitude of the stable limit cycle. An application of this method for a forced Van der Pol
generalized oscillator equation is made and the approximate solution found is valid for any order of the variable $t$.
The numerical simulation of the forced Van der Pol generalized oscillator equation on one hand and the
approximate solution found on the other hand for some values of control parameters show the method efficiency
and the validity of the approximate solution found. We have noticed that these control parameters play a key role
in the dynamic of the system. One notices also the primary resonance presence near the $\Omega= 1$ 
zone and that
the system presents a classical Hopf’s bifurcation through the renormalization equation.

{\bf{Acknowledgments}}

The authors thank IMSP for his  quality formation
and the beninese state which  has fully funded this work.

\end{document}